\documentclass[draft]{agujournal2019}
\usepackage{lineno}
\usepackage[inline]{trackchanges} 
\usepackage{soul}
\usepackage{amssymb}
\usepackage{amsmath}
\usepackage{bm}
\usepackage{float}
\usepackage{ragged2e}
\usepackage{float}
\usepackage{makecell}
\usepackage{xurl}
\usepackage{xcolor}
\justifying

\draftfalse

\journalname{Space Weather}

\begin{document}

%
%


\title{Forecasting High-Speed Solar Wind Streams from Solar Images}

%
%




\authors{Daniel Collin\affil{1,2}, Yuri Shprits\affil{1,3,4}, Stefan J. Hofmeister\affil{5}, Stefano Bianco\affil{1}, Guillermo Gallego\affil{2,6}}

\affiliation{1}{Space Physics and Space Weather, GFZ Helmholtz Centre for Geosciences, Potsdam, Germany}
\affiliation{2}{Department of Electrical Engineering and Computer Science, Technical University of Berlin, Berlin, Germany}
\affiliation{3}{Institute of Physics and Astronomy, University of Potsdam, Potsdam, Germany}
\affiliation{4}{Department of Earth, Planetary, and Space Sciences, University of California Los Angeles, Los Angeles, USA}
\affiliation{5}{Columbia Astrophysics Laboratory, Columbia University, New York, USA}
\affiliation{6}{Einstein Center Digital Future, Berlin, Germany}





\correspondingauthor{Daniel Collin}{collin@gfz.de}




\begin{keypoints}
\item We use polynomial regression to predict the solar wind speed at Earth from coronal holes in solar EUV images, with an RMSE of 68.1 km/s.
\item We find that the mean squared error loss underpredicts the high-speed stream peak speeds and fix this with a distribution transformation.
\item Using the coronal hole area, location, and 27-day persistence speed, we predict high-speed streams more accurately than neural networks.
\end{keypoints}

%
%

%
%


\begin{abstract}
The solar wind, a stream of charged particles originating from the Sun and transcending interplanetary space, poses risks to technology and astronauts. In this work, we develop a prediction model to forecast the solar wind speed at the Earth. We focuse on high-speed streams (HSSs) and their solar source regions, coronal holes. As input, we use the coronal hole area, extracted from solar extreme ultraviolet (EUV) images and mapped on a fixed grid, as well as the solar wind speed 27 days before. We use a polynomial regression model and a distribution transformation to predict the solar wind speed with a lead time of four days. Our forecast achieves a root mean square error (RMSE) of 68.1 km/s for the solar wind speed prediction and an RMSE of 76.8 km/s for the HSS peak velocity prediction for 2010 to 2019. We also demonstrate the applicability of our model to the current solar cycle 25 in an operational setting, resulting in an RMSE of 80.3 km/s and an HSS peak velocity RMSE of 92.2 km/s. The study shows that a small number of physical features explains most of the solar wind variation, and that focusing on these features with simple machine learning algorithms even outperforms current approaches based on deep neural networks and MHD simulations. In addition, we explain why the typically used loss function, the mean squared error, systematically underestimates the HSS peak velocities, aggravates operational space weather forecasts, and how a distribution transformation can resolve this issue. 
\end{abstract}

\section*{Plain Language Summary}
The Sun constantly releases charged particles, referred to as the solar wind, which can damage technology and pose risks to astronauts. Therefore, we need to predict the solar wind multiple days in advance, especially fast solar wind streams emitted by coronal holes. Currently, many models have difficulties to predict the peak velocity of these solar wind streams accurately. We explore this problem by developing a new prediction model to forecast the solar wind speed, focusing on coronal holes. The model uses solar images and solar wind measurements to predict the solar wind speed at the Earth four days in advance. We show how to overcome the problem of underestimating the peak velocities of solar wind streams by applying a statistical transformation to the predictions, making predictions more reliable. By testing the model on almost ten years of data, we find that it is more accurate than much more complex models, such as modern artificial intelligence models. We show that the size and location of coronal holes as well as past solar wind speed measurements are the most important features for these predictions. Additionally, we demonstrate that our model can be used for an operational real-time forecast on recent data.

\section{Introduction}

Space weather effects in the near-Earth space environment pose threats to technological infrastructure, both in space and on Earth. In particular geomagnetic storms, primarily caused by high-speed solar wind streams (HSSs) and coronal mass ejections (CMEs), can lead to severe damage. In this study, we focus on high-speed streams, which originate from coronal holes. These long-lasting regions on the Sun possess a reduced density and temperature, as compared to the surrounding corona, and are characterized by a magnetic field topology that is open towards interplanetary space.
Along these field lines, plasma is accelerated away from the rotating Sun, forming HSSs that transcend the heliosphere \cite{Krieger73}. 
The interaction of these fast solar wind streams with the preceding slower ambient plasma forms a compression region and sometimes a shock wave, which can cause disturbances in Earth's magnetosphere and initiate geomagnetic storms \cite{Tsurutani97}.
During the declining and minimum phase of the solar cycle, HSSs originating from coronal holes are the dominating cause of geomagnetic storms, while during solar maximum, CMEs are the major cause \cite{Richardson00,Tsurutani06}. To assess these risks, a reliable forecast algorithm for the solar wind speed (SWS) is essential.

Due to their reduced temperature, the source coronal holes of HSSs can be identified well in solar extreme ultraviolet (EUV) images, revealing properties like their area and location \cite{Krista09,Rotter12,Heinemann19}. 
The emitted solar wind usually takes between two and five days to arrive at Earth, where it is observed by satellites in the Lagrange 1 point (L1). Multiple studies have investigated the relationship between coronal holes and the measured SWS as this facilitates predicting HSSs with a lead time of four days \cite{Rotter15,Upendran20,Raju21,Brown22}.
\citeA{Nolte76} show that there is an approximately linear dependence between the area of near-equatorial coronal holes and the peak of the SWS of the associated HSS.
\citeA{Rotter15} use that fact and introduce a linear forecasting model for the SWS based on the coronal hole area.
\citeA{Hofmeister18} confirm that relationship, and find the co-latitude, i.e., the latitudinal separation angle between the coronal hole and Earth, as another influence.
Additionally, coronal holes evolve slowly and persist on the solar surface for long periods, some for as long as twelve solar rotations \cite{Bohlin76,Heinemann20}.
At each rotation, the associated HSS structures cross Earth, imprinting the periodicity of the solar rotation rate of 27 days onto the near-Earth SWS measurements \cite{Sheeley77,Sargent85,Diego10}.
\citeA{Owens13} exploit this property by introducing the 27-day persistence model, using the SWS observed 27 days ago as a forecast. This model serves as a widely adopted benchmark and the SWS persistence is used in many more complex data-driven SWS forecasts \cite{Yang18, Bailey21, Brown22, Sun22}.

In the past, SWS forecasting was done by empirical methods, e.g., the empirical WSA \cite{Arge03}, or the exploitation of the linear relationship between the coronal hole area and the SWS \cite{Rotter15}. Another approach to that problem uses magnetohydrodynamic (MHD) simulations, e.g., WSA-ENLIL \cite{Odstrcil03} or EUHFORIA \cite{Pomoell18}. Especially MHD models are computationally demanding, and none of the approaches utilizes the fast-growing amount of available satellite data. 
Because of that, purely data-driven machine learning approaches, and in particular deep learning models, are gaining attention \cite{Yang18,Upendran20,Bailey21,Raju21,Brown22}.

In this study, we combine machine learning with known empirical relationships between coronal holes and the solar wind speed to develop a prediction model that forecasts the SWS four days in advance. 
First, we create a dataset, based on a small number of physically meaningful input features,
Then, we develop a real-time forecast model based on a polynomial regression. It is capable of modeling nonlinearities and interactions between features, but still simple enough to keep the model fully explainable.
We show that by using this model and the coronal hole area, its location, the 27-day persistence speed, and the sunspot number as input, we can reconstruct the SWS variations well.
Further, we explain the systematic underestimation of HSS peak velocities, found by various previous studies, and show how applying a post-training distribution transformation can resolve this issue.
Thereafter, we show that our simple algorithm even outperforms sophisticated deep learning models, thus allowing a strong reduction of the model complexity and improving interpretability.
Finally, we demonstrate that our model is capable of providing reliable solar wind predictions in an operational setting in the current solar cycle 25.

The paper is structured as follows: Section 2 explains the machine learning model, including the input dataset, the prediction model, and the evaluation. Section 3 evaluates the approach on almost ten years of data and analyzes its impact on SWS forecasting. Section 4 compares our approach to other models and Section 5 sets up an operational forecast. Section 6 discusses the main findings and Section 7 draws conclusions for future research.

\section{Machine Learning Model}
\label{sec:ml_model}

\begin{figure}
\centering
\includegraphics[width=\textwidth]{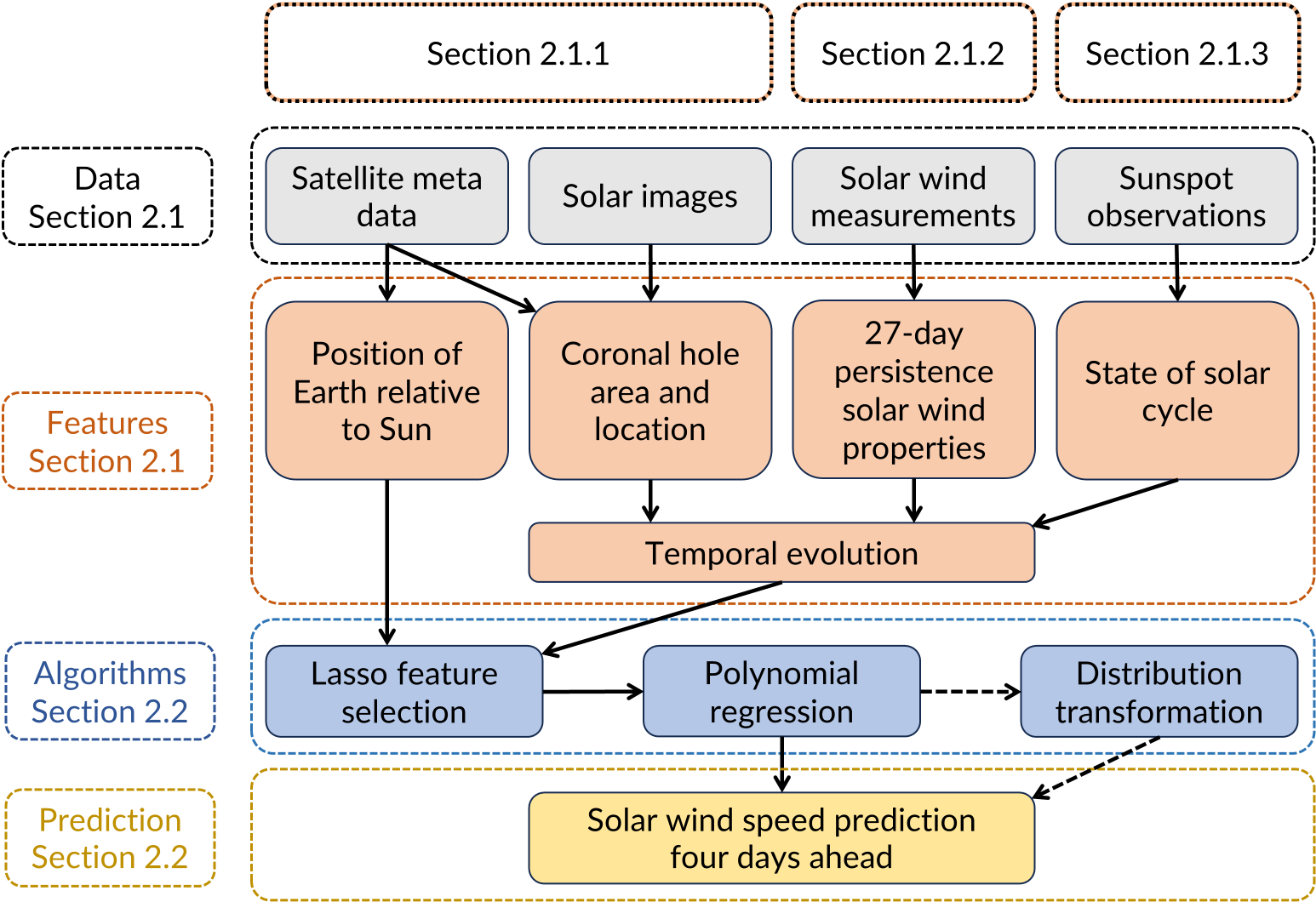}
\caption{Prediction Pipeline, indicating the corresponding sections of the paper. The dashed arrows indicate an optional path.}
\label{fig:pipeline}
\end{figure}

Our model is structured as depicted in Figure \ref{fig:pipeline}. 
Using real-time satellite data and sunspot observations, we compute input features containing information about the coronal holes that are currently visible on the solar surface, recurring solar wind streams emitted by them, and the current state of the solar cycle.
These features are then used to predict the SWS four days ahead, as for this forecast horizon there exists the highest cross-correlation between the total coronal hole area and the SWS. The forecast uses a polynomial regression as the main algorithm.

The main input in our dataset is the coronal hole area and location, which we extract from solar images using a coronal hole segmentation algorithm and a grid structure placed on the resulting segmentation maps
(Section \ref{sec:solar_images}).
From the satellite metadata, which specifies the position of the Earth relative to the Sun, we determine which solar hemisphere is tilted towards the Earth. To include solar wind persistence data, we use properties of the solar wind measured at L1 one solar rotation before
(Section \ref{sec:sw_measurements}).
To that we add the monthly sunspot number and its trend, provided by sunspot observations on the solar surface, as it is possible to estimate the solar cycle phases from it
(Section \ref{sec:ssn_observations}).
All physical parameters are sampled over multiple days to capture their temporal evolution, and used as input to the prediction model.

The prediction model is composed of three algorithms: First, we perform an automatic feature selection, using a linear model with Lasso regularization.
Features with negligible impact are discarded, which reduces the complexity of the model and increases the interpretability. In the second step, we train a polynomial regression model, which allows capturing the nonlinear interactions of the remaining features.
Third, we apply a distribution transformation to the output of the polynomial model to increase the accuracy of HSS peak velocity predictions.

In the following, we explain these steps in detail.

\subsection{Dataset}
\label{sec:dataset}

We divide the dataset into three categories: features extracted from solar images,
from solar wind measurements,
and from sunspot observations.
To differentiate between measurements recorded at multiple time points, we use the notation $x^{(t)}$ to denote a feature $x$ recorded $t$ days in the past relative to the forecast time. All features are listed in Table 1. Note that the total number of features depends on the resolution of the grid which is used for extracting the coronal hole area from the solar images. Datasets based on a 4$\times$3, 6$\times$6, or 10$\times$10 grid have 66, 162, or 418 features, respectively. The target output is the SWS at time $t=0$ days. 
We create two datasets. The first one covers solar cycle 24 and is used for model development and cross-validation. It spans the time range from June 2010 to December 2019 at a cadence of one hour and therefore comprises a total of 84024 data points. The second dataset covers the rising phase of solar cycle 25 and is used to evaluate the developed model in an operational setting. It spans the time range from January 2020 to December 2023 at a cadence of one hour and comprises of a total of 35064 data points.
From these datasets, we remove all CMEs using the Richardson \& Cane ICME list \cite{RichardsonCane,RCdata}, resulting in 45933 and 21064 remaining data points, respectively.

\begin{table}
\label{tab:features}
\caption{Input features used to predict the SWS at time $t=0$ days. $t$ denotes the number of days before the forecasted time point. ex. = extrapolated.}
\centering
\begin{tabular}{lll}
\hline
Feature & Definition & $t$ \\
 \hline
 $\alpha^{(t)}$ & Heliospheric latitude of Earth & 4\\
$S_{i,j}^{(t)}$ & \makecell[tl]{Coronal hole area of two solar grid\\cells symmetric to the equator} & 4,5,6,7\\
$D_{i,j}^{(t)}$ & \makecell[tl]{Asymmetry of coronal hole area\\between solar hemispheres } & 4,5,6,7 \\
$v^{(t)}$ & Solar wind speed &26,27,28 \\
$\rho^{(t)}$ & Solar wind density &26,27,28\\
$p^{(t)}$ & Solar wind pressure &26,27,28\\
$T^{(t)}$ & Solar wind temperature  &26,27,28\\
$B^{(t)}$ & Magnetic field magnitude & 26,27,28\\
$N_{\text{SS}}^{(t)}$ & Smoothed monthly sunspot number & 0 (ex.)\\
$\Delta N_{\text{SS}}^{(t)}$ & Change of smoothed monthly sunspot number & 0 (ex.)\\
\hline
\end{tabular}
\end{table}

\subsubsection{Solar Images}
\label{sec:solar_images}

To analyze the properties of coronal holes, we use solar EUV images taken by the Atmospheric Imaging Assembly (AIA) telescope onboard the Solar Dynamics Observatory (SDO) spacecraft \cite{Lemen12}.
Its 193 Å filtergrams show primarily the emission of Fe XII ions in the corona at temperatures of approximately 1.6 MK. Its 211 Å filtergrams show primarily the emission of Fe XIV ions at temperatures of approximately 2.0 MK. In both channels, coronal holes can be clearly seen.
To determine the position of Earth relative to the Sun, we take the spacecraft position in heliographic inertial coordinates of satellites at L1 from the OMNI\_M data provided by the NASA COHOWeb.

\begin{figure}
\centering
\includegraphics[width=\textwidth]{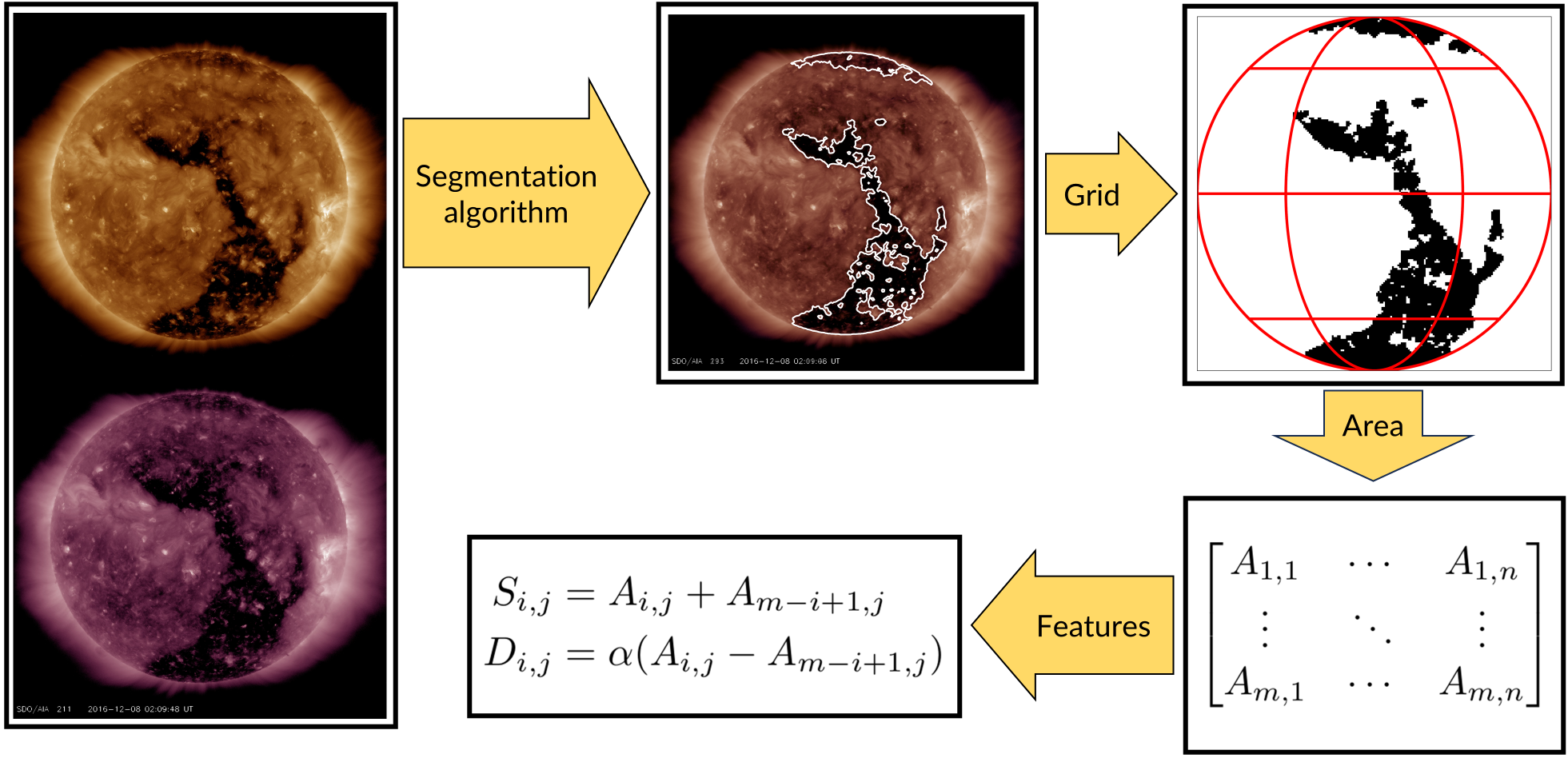}
\caption{Feature extraction from solar images. We detect coronal holes, place a grid structure (here, a 4$\times$3 grid) on the segmentation maps, extract the coronal hole area $A_{i,j}$ of each grid cell, and compute the features $S_{i,j}$ and $D_{i,j}$, quantifying the coronal hole area and its asymmetry between the hemispheres.}
\label{fig:preprocessing}
\end{figure}

From the solar images, we extract the coronal hole area and its location by using the segmentation algorithm of \citeA{Inceoglu22} (see Figure \ref{fig:preprocessing}).
The idea consists of combining the 193 and 211 Å channels into a two-channel image and applying a $k$-means clustering
to the pixel intensities. Choosing $k=3$, this method groups the pixels into three subsets with similar brightness, belonging to coronal holes, active regions, and the quiet Sun. For more details, we refer to \citeA{Inceoglu22}. The pixels assigned as coronal holes are translated to a binary map and downsampled to a map of 256$\times$256 pixels.
To exclude outliers, we compute the total coronal hole area from each map and remove those that exceed five times the median absolute deviation of a sliding window of ten observations (described as the Hampel filter in \citeA{Liu04}).
We replace these excluded outliers by interpolating between the adjacent time steps.

Then, we place an $m\times n$ grid on the coronal hole maps. It divides the solar disk into $m$ latitudinal and $n$ longitudinal cells of equal extent.
We use a variety of grid resolutions in this study, from a trivial 1$\times$1 grid up to a complex 14$\times$10 grid.
For each grid cell $(i,j)$
we extract the coronal hole area, which we denote as $A_{i,j}$. We do not need to correct for the curvature of the solar disk as the machine learning model learns the necessary scaling by itself.
Next, we add the heliospheric latitude of Earth $\alpha$, i.e., the angle between the solar equatorial plane and the current position of the Earth.
In case the Earth is located in the solar equatorial plane, i.e., $\alpha=0$, we assume that area sectors, which are symmetric to the solar equator, contribute equally to the solar wind stream. For $\alpha\neq0$, however, we expect the sector tilted towards the Earth to be more relevant than the other one tilted away.
Therefore, we introduce the weightings
$(1+ w_{i,j}\alpha)A_{i,j}$ (northern hemisphere) and $(1- w_{i,j}\alpha)A_{m-i+1,j}$ (southern hemisphere), where $w_{i,j}>0$ is a learnable scaling factor. 
By introducing $S_{i,j}:=A_{i,j}+A_{m-i+1,j}$ and $D_{i,j}:=\alpha (A_{i,j} - A_{m-i+1,j})$, we can rewrite the contribution of the area sectors as
\begin{equation*} 
    A_{i,j}+A_{m-i+1,j} + w_{i,j}\alpha (A_{i,j} - A_{m-i+1,j}) =  S_{i,j} + w_{i,j}D_{i,j}.
\end{equation*}
As the model automatically learns the scaling factor $w_{i,j}$, that results in two features: $S_{i,j}$, containing the coronal hole area, and $D_{i,j}$, quantifying the asymmetry between the northern and southern hemispheres.
To predict the SWS with a lead time of four days, we use the coronal hole features from four to seven days before the predicted time point at a cadence of one day as input features. That allows us to capture the temporal evolution of the coronal holes.

\subsubsection{Solar Wind Measurements}
\label{sec:sw_measurements}

To obtain solar wind properties at the Earth, we use hourly averages of in-situ plasma measurements recorded at L1 as provided by the NASA OMNIWeb database \cite{OMNI}. 
OMNIWeb provides these measurements time-shifted to the Earth's bow shock to be aligned with measurements of other spacecraft closer to Earth. 
We capture the periodicity of the solar wind structure of recurring coronal holes by incorporating solar wind persistence data from one solar rotation ago.
We include the solar wind speed $v$, solar wind density $\rho$, solar wind pressure $p$, solar wind temperature $T$, and the magnitude of the average magnetic field vector $B$, all sampled from 26, 27, and 28 days before the predicted time point.

\subsubsection{Sunspot Observations}
\label{sec:ssn_observations}

As an indicator of the current state of the solar cycle, we incorporate the monthly sunspot number from the NOAA Space Weather Prediction Center. That value can be used to estimate the current state of the solar cycle by considering both the current sunspot number and its rate of change.
The sunspot number is provided every month and thus not available for the current date in a real-time forecast. Therefore, we extrapolate it. 

First, we denoise the time series by fitting a quadratic function to the last 48 monthly values. That provides a smoothed sunspot number curve and its slope. 
As changing trends cannot be anticipated well, the extrapolation of a time series is prone to generating a curve that lags behind the observed curve by a couple of data points. We find that this lag can be mitigated by extrapolating the monthly sunspot number two months after the forecasted time point and using it as the extrapolated value. To decrease oscillations of the extrapolated curve and to increase the robustness against outliers, we use a linear extrapolation based on the function value and slope six months before the forecasted time point.
With this, we obtain a smoothed monthly sunspot number $N_{\text{SS}}$, describing the current level of solar activity. The slope produces another feature $\Delta N_{\text{SS}}$, describing the current rate of change of the smoothed sunspot number and indicating if solar activity is rising or decreasing. To obtain hourly values as needed for our dataset, we interpolate between the monthly values.

\subsubsection{High-Speed Streams and Coronal Mass Ejections}
\label{sec:hss_cme}

Our model focuses on HSSs related to coronal holes and does not take into account CMEs.
Therefore, their occurrence cannot be predicted and we exclude all CME-related disturbances from our data using the Richardson \& Cane ICME list \cite{RichardsonCane,RCdata}. Additionally, we aim to evaluate our model's prediction accuracy for properties of HSSs. Thus, we need to filter all SWS enhancements, i.e., extended periods of above-average solar wind velocity, from the solar wind time series and divide them into a set of HSSs and CME-related disturbances.

First, we use a simple peak finding algorithm on the denoised time series to define SWS enhancements. The denoising is performed by applying a Gaussian filter with a standard deviation of 24 hours.
Then, all peaks that are at least four days apart, exceed 390 km/s, and possess a prominence, i.e., vertical distance between its highest point and its base, of at least 35 km/s, are considered as enhancements. We define the start and end of the enhancement at the points where the smoothed time series crosses the relative height of 0.4 from the peak's base to its maximum. In case that two enhancements overlap, we truncate the longer one.

Then, we match predicted and observed enhancements, following the approach from \citeA{Reiss16} with slight changes. 
The predicted and observed peaks are associated if they are within three days of each other. If several enhancements meet this condition, we match the temporally closest ones and mark them as a hit. 
The remaining unmatched predicted enhancements are labeled as false alarms, and the remaining observed enhancements are labeled as misses. 

Finally, we ensure that the influence of CMEs is excluded from our study. We classify as a CME disturbance all ICMEs from the list of Richardson \& Cane and all observed SWS enhancements that intersect with an ICME or happen until two days afterwards. All other enhancements are labeled as an HSSs.
Then, we remove all intervals of CME disturbances and, since we also use 27-day persistent features, all data points from one solar rotation thereafter from the entire machine learning dataset and from our separate list of SWS enhancements.

In summary, we obtain two curated datasets. First, a machine learning dataset of solar wind features with all effects of CMEs on our dataset excluded. Second, a list of observed and predicted HSS events.

\subsection{Prediction Model}
\label{sec:prediction_model}

In the following, we describe the feature selection,
the polynomial regression model,
and the distribution transformation
we use for the forecast model.

\subsubsection{Feature Selection}
\label{sec:fs}

Before training our model, we scale all features and the target to $[0,1]$ by min-max normalization. Then, we use a linear regression model regularized with the Least Absolute Shrinkage and Selection Operator (Lasso) for feature selection. The Lasso regularization adds to the least squares loss function of the regression a penalty term $\gamma ||\beta||_1$, where $\beta$ is the vector of model coefficients and $||.||_1$ is the $L^1$ norm. This term enforces sparsity in the learned coefficients, i.e., sets fitted model coefficients to zero if they have negligible impact on the output. Additionally, it thereby prevents overfitting and reduces the model complexity. The strength of the regularization is controlled by the hyperparameter $\gamma>0$ \cite{Tibshirani96}. The values of $\gamma$ are determined by a 5-fold cross-validation as described in Section \ref{sec:cv} and are given in \ref{app:grid}.

After scaling the input and fitting the linear Lasso model, the largest learned coefficient values of the linear feature selection model are in the order of $10^{-1}$. Thus, all features with absolute coefficient values below $10^{-4}$ have almost no impact on the final prediction and are deleted from the input to reduce the dimension of our dataset.

\subsubsection{Polynomial Regression}
\label{sec:pr}

We use a polynomial regression as the main machine learning algorithm. This model extends a regular linear regression model by incorporating higher powers and multiplicative interactions of features up to a specified degree, in order to capture nonlinear effects.
After the feature selection, the reduced feature set is used to fit a polynomial regression model of order 3, again regularized with the Lasso penalty. The values of the penalty parameter $\gamma$ are also shown in \ref{app:grid}.

Input features may take on values outside of the training data domain, for example, if unseen coronal hole distributions arise in the test dataset.
However, polynomial functions can exhibit steep gradients beyond the fitted intervals.
To mitigate the risk of predicting unrealistically strong downward variations of the SWS, we determine the minimum observed velocity in the training data and set all predictions below that threshold to this value. For upward variations, this issue is less problematic because it only leads to an extrapolation of the SWS peaks to higher velocities. 

\subsubsection{Distribution Transformation}
\label{sec:dt}

Machine learning models often tend to underestimate extreme events due to their underrepresentation in the training data. We investigate if this bias can be corrected by applying a distribution transformation that maps the distribution of the output of the machine learning model onto the distribution of observations. Analogous methods, called histogram matching, exist in the field of image processing. For this purpose, we use a Box-Cox transformation \cite{BoxCox}. 

The Box-Cox transformation is a statistical technique used to stabilize the variance of a dataset and to make the data more normally distributed. For a data point $y$, the transformation is defined as
\begin{align*}
    y^{(\lambda)}:=F_\lambda(y)
    =
        \begin{cases}
        \frac{{y^\lambda - 1}}{\lambda}, & \text{if } \lambda \neq 0 \\
        \log(y), & \text{if } \lambda = 0
        \end{cases},
        \quad \lambda \in\mathbb{R},
\end{align*}
where $\lambda$ needs to be fitted by maximizing the log-likelihood of the transformed dataset, i.e., maximizing the probability of observing the given data under the assumption that the transformed data follows a normal distribution.
The transformation can be used to map a distribution of data points onto another one. Given two datasets $\mathcal{A}$ and $\mathcal{B}$, e.g., predictions and observations, we fit $\lambda$ for each of them, apply the Box-Cox transformation, and normalize the transformed distributions by scaling them to zero mean and unit variance. Thereafter, the transformed and scaled distributions $\widetilde F_{\lambda_1}(\mathcal{A})$ and $\widetilde F_{\lambda_2}(\mathcal{B})$ approximately coincide. To further transform $\widetilde F_{\lambda_1}(\mathcal{A})$ into the desired original distribution $\mathcal{B}$, we apply the inverse of the normalization and the inverse of the Box-Cox transformation that were fitted to $\mathcal{B}$:
\begin{equation*}
    \mathcal{A}\xrightarrow[\text{transform}]{\text{fit }\lambda_1}F_{\lambda_1}(\mathcal{A}) \xrightarrow[]{\text{normalize}}\widetilde F_{\lambda_1}(\mathcal{A})\approx \widetilde F_{\lambda_2}(\mathcal{B})\xrightarrow[\text{normalize}]{\text{inverse}}F_{\lambda_2}(\mathcal{B})\xrightarrow[\text{transform with }\lambda_2]{\text{ inverse}} \mathcal{B}
\end{equation*}
By linking these operations, we define a fixed mapping that directly maps distribution $\mathcal{A}$ onto distribution $\mathcal{B}$.

To improve the prediction accuracy of HSS events, we apply this distribution transformation to the output of the polynomial model. This mapping is fitted on the training data and applied as postprocessing to the predictions. Its effect will be further discussed in Section \ref{sec:dt_analysis}.

\subsection{Evaluation}
\label{sec:evaluation}

In the following, we explain the evaluation of our approach. 
First, we introduce the used metrics,
and then, we explain our cross-validation scheme.

\subsubsection{Metrics}
\label{sec:metrics}

We evaluate the accuracy of our model with the root mean square error (RMSE), the mean absolute error (MAE), and the Pearson correlation coefficient (CC), which are standard metrics quantifying the point-to-point errors between the predictions and observations. If we compute these metrics on the whole continuous time series, we call them timeline RMSE, timeline MAE, and timeline CC.

Additionally, we use an event-based evaluation. Following the procedure outlined in Section \ref{sec:hss_cme}, we match observed and predicted HSSs and label all events as hit (correct HSS prediction), miss (HSS observation not predicted), or false alarm (HSS prediction not observed).
We count the number of hits as true positives (TP), the number of misses as false negatives (FN), and the number of false alarms as false positives (FP). Based on these values, the verification measures probability of detection (POD), false alarm ratio (FAR), threat score (TS), and bias (BS) can be calculated \cite{Woodcock76}: 
\begin{equation*}
     \text{POD}=\frac{\rm TP}{\rm TP + \rm FN},\quad \text{FAR}=\frac{\rm FP}{\rm TP + \rm FP},\quad \text{TS}=\frac{\rm TP}{\rm TP + \rm FP + \rm FN}, \quad  \text{BS}=\frac{\rm TP + \rm FP}{\rm TP + \rm FN}.
\end{equation*}
The POD quantifies the HSS detection ratio among all observed HSSs, the FAR the ratio of false alarms among all predicted HSSs, and the TS combines both aspects for the overall model performance. These metrics range between 0 and 1. For the POD and the TS, the best possible outcome is 1 and the worst one is 0. For the FAR, 0 is the best and 1 is the worst outcome. A BS smaller than 1 indicates that the number of events is underestimated and a BS greater than 1 that the number of events is overestimated. A BS of 1 implies an unbiased prediction.

Lastly, we use the matching of observed and predicted HSSs to apply the previously defined point-to-point metrics to the HSS peak velocity predictions to quantify the HSS peak velocity RMSE, HSS peak velocity MAE, and HSS peak velocity CC of associated solar wind streams.

\subsubsection{Cross-Validation}
\label{sec:cv}

\begin{figure}[H]
\centering
\includegraphics[width=0.6\textwidth]{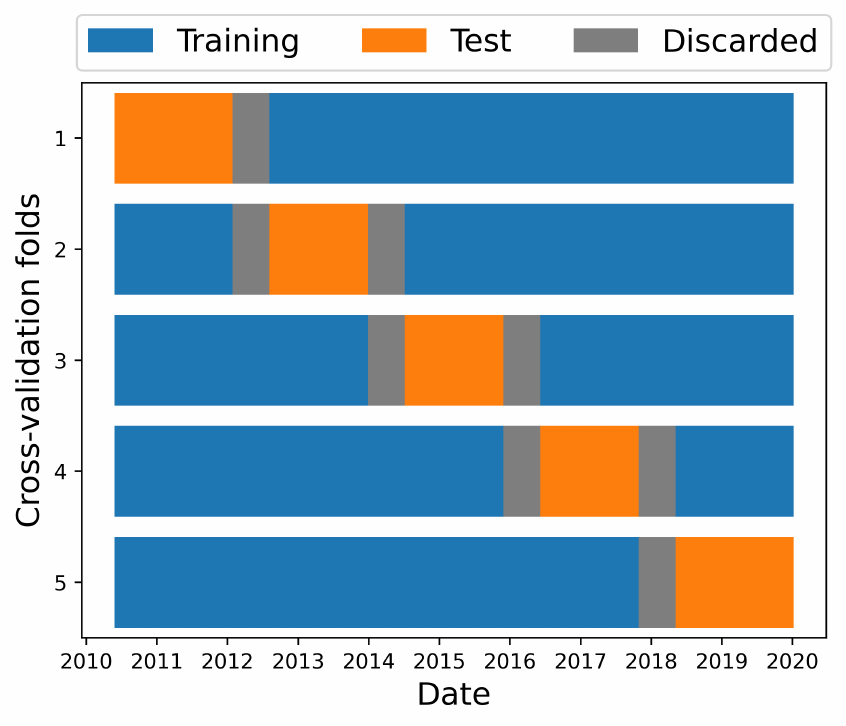}
\caption{Cross-validation data splits of our dataset. 180 days of data are discarded between training and test data.}
\label{fig:cv}
\end{figure}

We use 5-fold cross-validation (CV) to evaluate our model, i.e., we divide the dataset into five subsets, assign four of them as training data to fit the model, and one as test data to evaluate the model (see Figure \ref{fig:cv}). Iteratively, each pair of training and test data is used once to fit and evaluate the model. The model's generalization capability is assessed by computing the evaluation metrics on the test data.

The SWS time series exhibits an auto-correlation for up to 4 days and recurring auto-correlation peaks every multiple of 27 days, caused by long-lasting coronal holes that re-occur at the visible solar disk with each solar rotation. It is therefore crucial to discard a sufficiently long period of data between training and test datasets to avoid data leakage caused by these recurring coronal holes holes appearing in both training and test data. Thus, we divide the dataset into five subsets of approximately one year and eleven months each. We train the model on four subsets and evaluate it on the fifth subset, discarding 180 days of data at the interface of the training and evaluation subsets to avoid data leakage from long-lived coronal holes.
Figure \ref{fig:cv} shows the resulting CV datasets. Additionally, we exclude all CME disturbances from our training and test data as described in Section \ref{sec:hss_cme}. That results in a CV split with 37363, 37973, 37737, 31578, and 30562 training data points and 7146, 5081, 4947, 8751, and 11543 test data points, respectively.

To optimize the $\gamma$ hyperparameters of the Lasso regularization, we compare the model performance over all CV splits for different sets of $\gamma$. We employ the tree-structured Parzen estimator (TPE) approach, a greedy, sequential method based on the expected improvement criterion, to find the best $\gamma$ hyperparameters \cite{Bergstra11,Bergstra13}.

\section{Results and Analysis}

In the following, we first discuss which features result in the best SWS predictions, aiming to find the polynomial model that fits the observations best by comparing various grid resolutions (Section \ref{sec:grid}) and analyzing the impact of each feature (Section \ref{sec:feature_importance}).
Then, we investigate the effect of additionally applying the distribution transformation to the predictions (Section \ref{sec:dt_analysis}), as well as the accuracy of our model (Section \ref{sec:solar_cycle}).

\subsection{Grid Resolution}
\label{sec:grid}

First, we focus on the grid that is used to bin the coronal hole area on the surface of the Sun. We analyze the effect that the grid resolution has on the capacity of the polynomial model to fit the target data by testing different $m\times n$ grids, where $m$ is the number of latitudinal and $n$ is the number of longitudinal cells. We start from the simplest grid possible, a 1$\times$1 grid, and compare a variety of grid resolutions up to a 14$\times$10 grid. Increasing the resolution further leads to grid cells becoming too small, just consisting of a couple of pixels, and hence a high likelihood of overfitting. We optimize the performance of each grid with respect to the timeline RMSE and the HSS peak velocity RMSE by doing a hyperparameter search. 
The values of the Lasso penalty parameters $\gamma$ are adjusted for each specific grid resolution and shown in \ref{app:grid}. 
We record the input parameters to the polynomial model after the feature selection step over all CV folds to infer the number of used features as a measure for the model complexity.

\begin{figure}
\centering
\includegraphics[width=\textwidth]{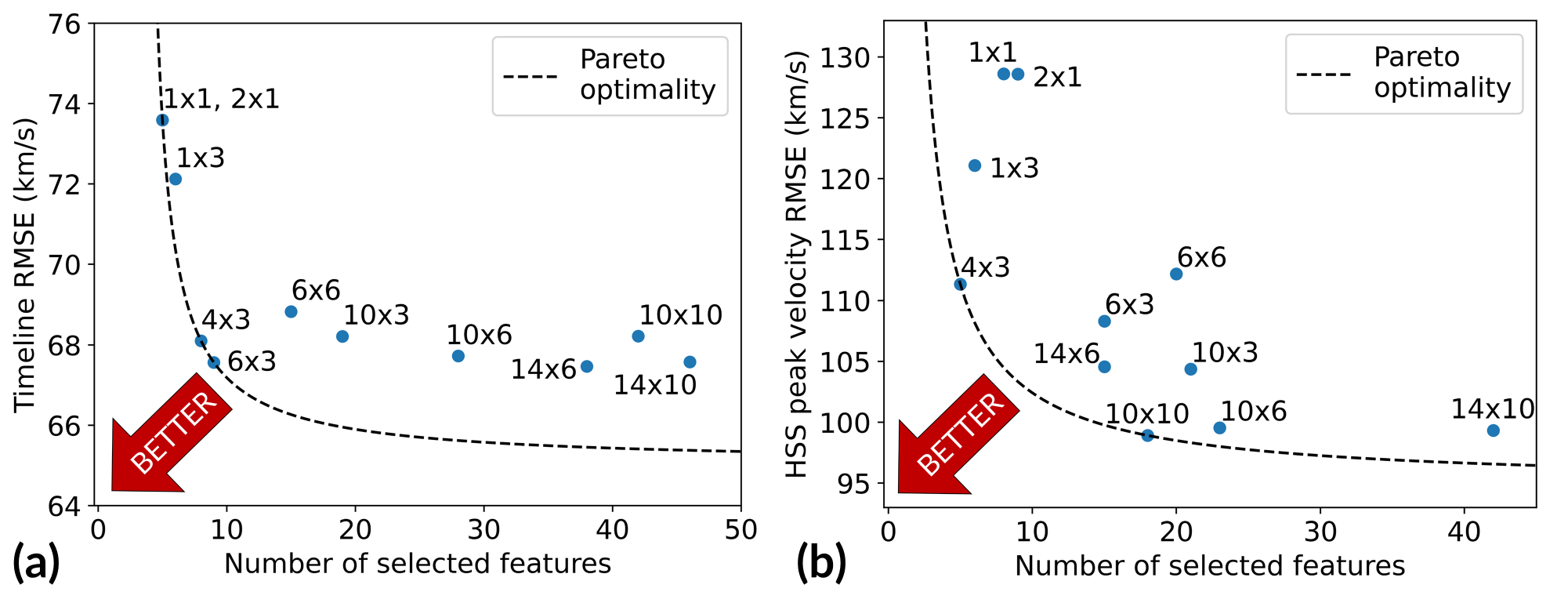}
\caption{Trade-off between performance and model complexity (number of features remaining after the feature selection) for models based on different grid resolutions. The dashed line is the Pareto front (see \ref{app:pareto}).} (a) Models whose hyperparameters were optimized to minimize the timeline RMSE. (b) Models whose hyperparameters were optimized to minimize the HSS peak velocity RMSE.
\label{fig:grid_comp}
\end{figure}

The results are presented in Figure \ref{fig:grid_comp}, showing the trade-off between accuracy and model complexity, quantified by the number of features selected by the feature selection model. A table of all grids, metrics, and hyperparameters is given in \ref{app:grid}. We search for the best-performing model with the least complexity. With respect to the timeline RMSE, we see that the medium resolutions of the 4$\times$3 and 6$\times$3 grid model achieve the highest accuracy while using a small number of features. Higher resolutions provide no further improvement, and lower ones restrict the model's expressivity too much to fit the data well. The number of features selected by the feature selection model increases with the grid size, which is plausible as the grid cell size shrinks and more cells need to be selected to capture the same surface area of the Sun. With respect to the HSS peak velocity RMSE, the 10$\times$10 grid model clearly outperforms all other grid choices. A finer grid in the latitudinal direction is advantageous, as there is an improvement in performance for all models with ten or more latitudinal cells. The peak accuracy of these models shows a further improvement if the number of longitudinal ranges increases. 

We conclude that a fine distinction of the location of coronal holes is crucial for predicting the maximum speed during an HSS correctly, but less important for the timeline accuracy. Additionally, the difference between the best and worst model for the timeline prediction is only 6 km/s, almost negligible for practical applications, whereas, in contrast, the difference for HSS peak predictions is 30 km/s, demonstrating significant variation. 
Therefore, using a grid-based approach is particularly beneficial for HSS peak velocity predictions. 
This can be attributed to the fact that the timeline predictions are dominated by the slow solar wind, which is unaffected by coronal holes, whereas HSS peak predictions are based solely on data points where the coronal hole distribution on the solar disk is relevant.
For the following analysis, we choose for a timeline model the 4$\times$3 grid.
Although its RMSE of 68.1 km/s is slightly higher than the RMSE of 67.6 km/s for the 6x3 model, it detects HSSs more reliably with a POD of 0.73 and a TS of 0.66, compared to the POD of 0.70 and the TS of 0.62 of the 6$\times$3 grid model. Additionally, it uses only eight features.
For peak velocity predictions, we choose the 10$\times$10 grid model. It uses 18 features to achieve an HSS peak velocity RMSE of 98.9 km/s, a POD of 0.77, and a TS of 0.69. 
Further, we notice for the metrics displayed in \ref{app:grid} that improving the performance at the HSS peaks decreases the timeline performance, and vice versa. This trade-off is confirmed in the subsequent experiments.

\subsection{Feature Importance}
\label{sec:feature_importance}

\begin{figure}
\centering
\includegraphics[width=\textwidth]{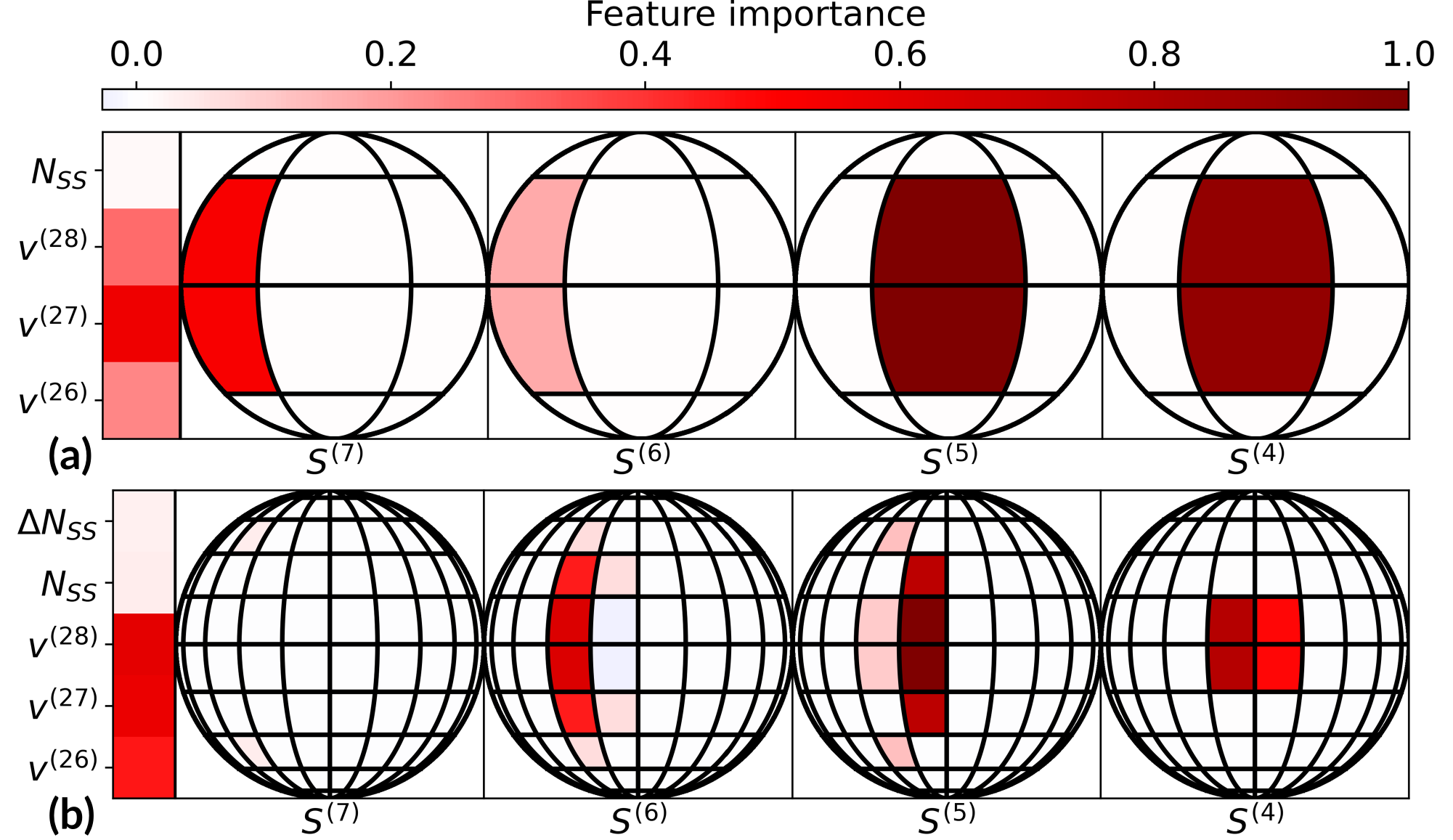}
\caption{Feature importance of the input features to the polynomial model. The coronal hole area $S^{(t)}$, the SWS $v^{(t)}$, both $t$ days before the prediction, the sunspot number $N_{\rm SS}$, and its change $\Delta N_{\rm SS}$ are the only parameters remaining after the feature selection. 
(a) Permutation feature importance with respect to the timeline RMSE for the 4$\times$3 grid model. 
(b) Permutation feature importance with respect to the HSS peak velocity RMSE for the 10$\times$10 grid model.}
\label{fig:importance}
\end{figure}

We perform a feature importance analysis of the polynomial model to study the relevance of the input features for the SWS prediction. The relevant set of features extracted by the feature selection mechanism is visualized in Figure \ref{fig:importance}. It is notable that in both the 4$\times$3 and the 10$\times$10 grid model only coronal hole area features $S_{i,j}^{(t)}$, the SWS $v^{(t)}$, as measured one solar rotation ago, and the sunspot number $N_{\text{SS}}$ and its change $\Delta N_{\text{SS}}$ remain and are used in the polynomial regression. Although other parameters like the heliospheric latitude of the Earth $\alpha$, which is also part of the $D_{i,j}$ features, might contribute to the HSS peak velocities, their influence on the timeline error is negligible. The reason is that the peaks are only a minor subset of the full training dataset. Since the feature selection is based on fitting a linear model with the timeline MSE loss function, it only means that their linear contribution to the overall time series prediction is negligible.

To assess the impact of the selected features on the RMSE of the polynomial model, we employ permutation feature importance \cite{Altmann10}. Given a dataset, it consists of choosing a feature and randomly permuting (shuffling) all of its values in the dataset. Thereafter, the permuted dataset is used as input to the trained model to compute new predictions. The original and new predictions are evaluated and the impact on a performance metric is observed.
If a feature is important, shuffling its values should significantly degrade the model's performance. We calculate the performance difference and scale the obtained values of all features to the interval $[0,1]$. For the HSS peak velocity RMSE, we accordingly evaluate how much the velocities change at the time points that correspond to the originally predicted velocity peaks.

We compute feature importance scores for the 4$\times$3 grid model with respect to the timeline RMSE and for the 10$\times$10 grid model with respect to the HSS peak velocity RMSE. The results are shown in Figure \ref{fig:importance}. The most important features for the timeline RMSE are the centrally located coronal hole areas four and five days ago. Then, the focus of the model shifts to the east as we move back in time. This is consistent with physical models, as coronal holes first become visible on the left, i.e., eastern side of the Sun, before rotating to the center, from where the solar wind stream is emitted towards Earth. The central coronal hole area four to five days in the past is therefore most relevant for the SWS, but the model also extracts useful information from the temporal evolution of the appearance of the coronal holes.
Higher latitudes are not taken into account, meaning that they do not provide a benefit in terms of prediction accuracy for the studied time period. 
The SWS observed between 26 and 28 days ago, particularly $v^{(27)}$, has a strong impact, aligning with physical knowledge due to the long lifetime of coronal holes and the velocity of the slow solar wind for the timeline evaluation.

For the HSS peak velocity RMSE, centrally located coronal holes also have the highest relevance, but it is advantageous to differentiate between latitudes. The importance decreases for greater distances from the solar equator. This can be explained by the fact that for HSSs arising from coronal holes at higher solar latitudes, Earth is farther in the flanks of the HSS, which results in lower HSS peak velocities measured \cite{Hofmeister18}.
It also explains why the peak accuracy improves when the latitudinal resolution of the grid is increased (shown in Section \ref{sec:grid}). 
Again, high latitudes are ignored, and the focus of the model shifts to the east as we consider area features recorded earlier.
The earliest time step, seven days in the past, is neglected, because typically, coronal holes close to the eastern and western limb can be barely seen due to overshining, i.e., line-of-sight integration of quiet-Sun features in line with the coronal holes. That leads to a deterioration of the coronal hole segmentation and less informative value for the forecast.

Additionally, there is one feature with a negative importance, which is $S_{5,5}^{(6)}$, meaning that randomly replacing the feature values improves the performance. That is a typical sign of overfitting on the training data, which then fails to generalize to the test data. The SWS from one solar rotation ago has a high importance, indicating that recurring HSSs have highly correlated peak velocities and these features are an important baseline for the prediction. The sunspot number $N_{\text{SS}}$ and its change $\Delta N_{\text{SS}}$ only play a minor role for both metrics, which might be attributed to the fact that only one solar cycle of data was included, preventing an adaption of the model to the systematic differences between the phases of the solar cycle. Furthermore, the other features have, relatively seen, a much more direct influence on the observed SWS than the variations of the solar cycle.

We conclude that a large portion of the solar wind variability can be explained by focusing on the near-equatorial coronal hole area and the SWS observed 27 days ago. Additionally, it is important to capture several timesteps of the observed coronal hole and solar wind time series.

\subsection{Distribution Transformation}
\label{sec:dt_analysis}

\begin{figure}
\centering
\includegraphics[width=0.97\textwidth]{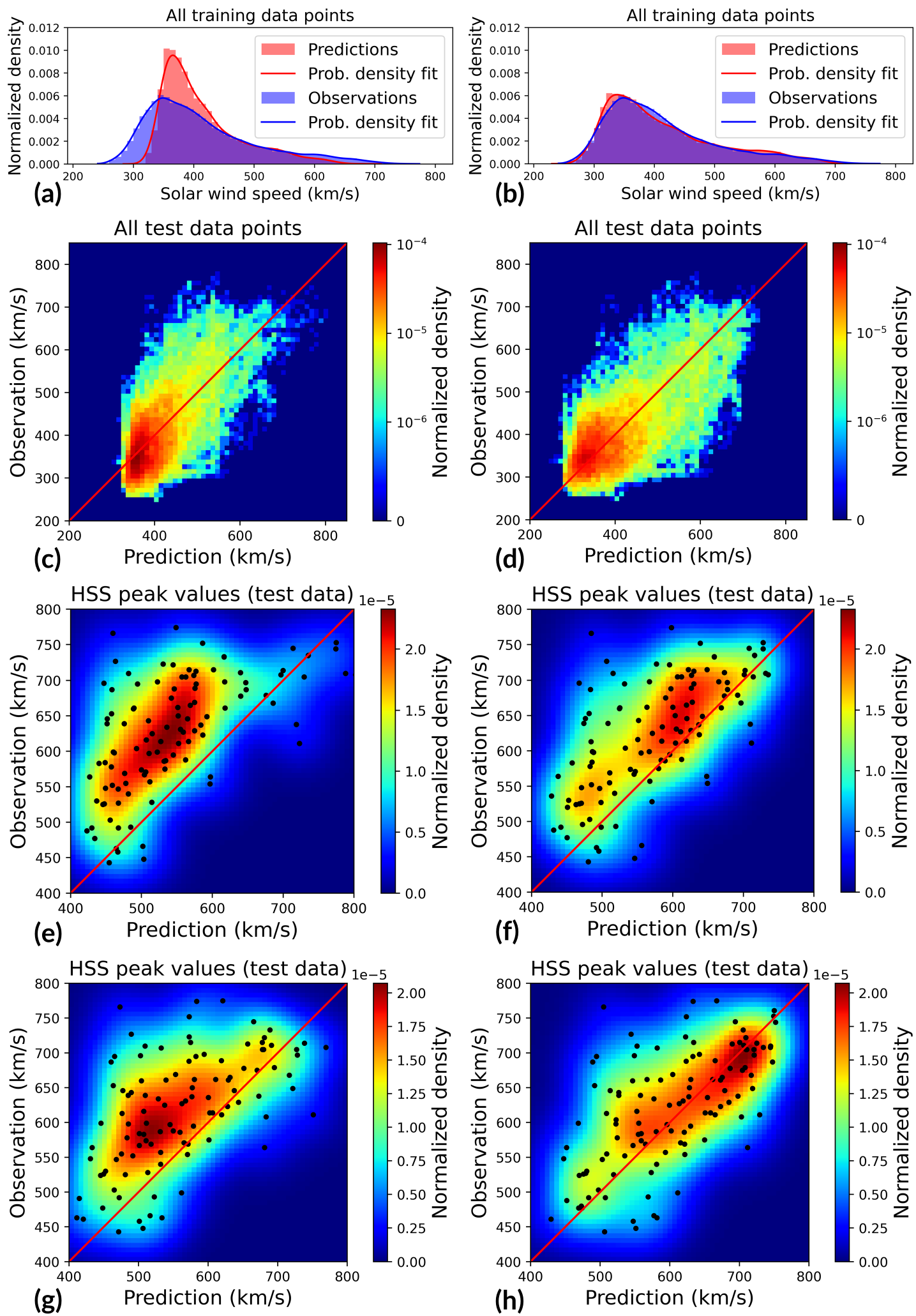}
\caption{(a) Histogram of observations and predictions
using the 4$\times$3 grid polynomial model. The bin size is 10 km/s.
(b) Same as (a) after distribution transformation (DT).
(c) Density plot of predictions vs. observations
using the 4$\times$3 grid model. The bin size is 11 km/s.
(d) Same as (c) after DT.
(e) Density plot of HSS peak velocity predictions vs. observations,
using the 4$\times$3 grid model; black dots show individual peaks, and the histogram is smoothed to approximate the density.
(f) Same as (e) after DT.
(g) Same as (e) for the 10$\times$10 grid model.
(h) Same as (g) after DT.
}
\label{fig:distr_trans}
\end{figure}

We analyze the effect of the distribution transformation by comparing the predictions before and after applying the transformation. For that purpose, we merge all predictions of the CV models on their respective training and test sets into one training and one test set. Figure \ref{fig:distr_trans} shows the effect on the distribution of SWS predictions. 
The histogram of training data distributions demonstrates that the polynomial model poorly fits the distribution of observations, neglecting the heavy tails and failing to predict slow speeds (Figure \ref{fig:distr_trans}a). 
This is particularly problematic for the peak velocities of HSSs, by definition consisting of speeds at the upper end of the observed interval. The distribution transformation accounts for that systematic error and scales the predictions
to the observed velocity distribution of the training data (Figure \ref{fig:distr_trans}b). 
The implications of the poor distribution fit can be observed also in the density plot of the predictions vs. observations.
For the predictions, there is a bias to underestimate the occurrence of high and low speeds, shown by the asymmetry around the identity line (Figure \ref{fig:distr_trans}c). This effect is removed by the distribution transformation, and the errors are spread symmetrically (Figure \ref{fig:distr_trans}d). That bias is even more clearly shown for the distribution of the associated HSS peak velocities. The density plots show how the majority of peak velocities is strongly underestimated, although the effect is slightly less pronounced for the 10$\times$10 grid model (Figures \ref{fig:distr_trans}e and g). If we apply the distribution transformation, we significantly mitigate this issue. Peak velocity predictions are scaled up and agree much better with the observed velocities and the 10$\times$10 grid model even achieves unbiased predictions (Figures \ref{fig:distr_trans}f and h).

The evaluation metrics before and after the distribution transformation are shown in Table \ref{tab:cont_metrics}. 
Regarding the detection rate of HSSs, the distribution transformation has a positive effect. We find that out of 147 observed HSS, the 4$\times$3 grid polynomial model without the distribution transformation detects 107, resulting in a POD of 0.73. These numbers increase to 113 and 0.77, respectively, when the distribution transformation is applied. The POD of the 10$\times$10 grid model even increases from 0.77 to 0.80, as it detects 113 HSSs before and 118 HSSs after the distribution transformation. The FAR slightly grows for both models, because all falsely predicted variations of the SWS are scaled as well, and some are then classified as HSSs. The combined effect of POD and FAR is positive for both models, as the 4$\times$3 grid model increases the TS from 0.66 to 0.69, and the 10$\times$10 grid model from 0.69 to 0.70. The BS of both models is getting closer to one, which means that there is a smaller bias to underestimate the occurrence of HSSs.

\begin{figure}
\centering
\includegraphics[width=0.6\textwidth]{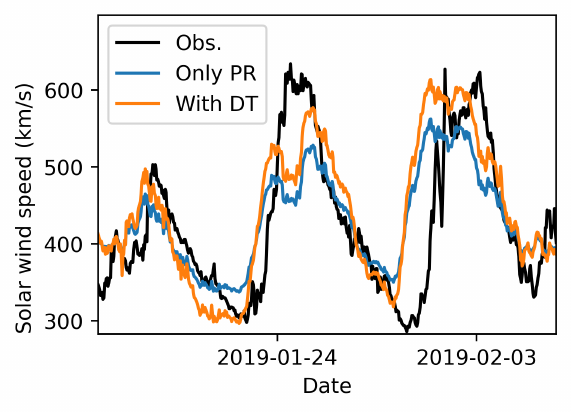}
\caption{Timeline prediction of the 10$\times$10 grid model, as it is output by the polynomial regression model (PR), and after additionally applying the distribution transformation (DT).}
\label{fig:trans_comp}
\end{figure}

The improvement of the predictions due to the distribution transformation can also be seen for the predicted velocity timeline, visualized in Figure \ref{fig:trans_comp}. We see that the scaling improves the quality of HSS predictions by adjusting the predicted velocity maxima and minima to approximate the ones observed during the HSSs and the slow solar wind, respectively. However, we can also see that there is a general temporal uncertainty of HSS predictions, exemplarily seen here as a slight positive shift of the velocity enhancement, which is not affected by the transformation.

\begin{table}
\caption{\label{tab:cont_metrics}Evaluation metrics for the 4$\times$3 and 10$\times$10 grid model before and after applying the distribution transformation to the output of the polynomial model. RMSE and MAE are given in km/s. Bold numbers indicate the best values.}
\begin{tabular}{l|lll|lll|llll}
\hline
 & \multicolumn{3}{c|}{Timeline} & \multicolumn{3}{c|}{HSS peak velocities} & \multicolumn{4}{c}{HSS events}\\
\hline
Model  &RMSE   & MAE &   CC   & RMSE  & MAE & CC     & POD           & FAR           & TS    & BS  \\
 \hline
 \multicolumn{11}{l}{\textbf{Polynomial model}}\\
4$\times$3 & \textbf{68.1} & \textbf{52.6} & \textbf{0.70} & 113.3 & 94.1 & 0.58 & 0.73 & \textbf{0.12} & 0.66 & 0.82 \\
10$\times$10 & 72.1 & 54.3 & 0.66 & 98.9 & 79.2 & 0.60 & 0.77 & 0.13 & 0.69 & 0.88 \\
\hline
\multicolumn{11}{l}{\textbf{With distribution transformation}}\\
4$\times$3  & 75.1 & 57.8 & \textbf{0.70} & 87.5 & 66.8 & \textbf{0.62} & 0.77 & 0.13 & 0.69 & 0.88 \\
10$\times$10 & 79.0 & 59.8 & 0.68 &  \textbf{76.8} & \textbf{58.2} & \textbf{0.62} & \textbf{0.80} & 0.16 & \textbf{0.70} & \textbf{0.95}\\
\hline
\end{tabular}
\end{table}

These findings are confirmed by the evaluation metrics in Table \ref{tab:cont_metrics}.
Both models improve all HSS peak velocity metrics. In particular, the 4$\times$3 grid model improves the HSS peak velocity RMSE by 23\% from 113.3 to 87.5 km/s, and the 10$\times$10 grid model by 22\% from 98.9 to 76.8 km/s. 
However, that improvement is at the expense of 
timeline predictions, where
the RMSE increases from 68.1 to 75.1 km/s for the 4$\times$3 grid model and from 72.1 to 79.0 km/s for the 10$\times$10 grid model.
However, the CC stays about constant at 0.70 for the 4$\times$3 grid model and increases from 0.66 to 0.68 for the 10$\times$10 grid model. 
Generally, we observe again the trade-off in the performance between predicting the HSS peaks and the timeline.

We conclude that the distribution transformation significantly improves the quality of the HSS peak velocity predictions and the HSS detection capacity. It overcomes the systematic bias of underestimating the peak velocities of HSSs.
However, it also slightly degrades the metrics of the timeline predictions.

\subsection{Solar Cycle}
\label{sec:solar_cycle}

\begin{figure}
\centering
\includegraphics[width=1.0\textwidth]{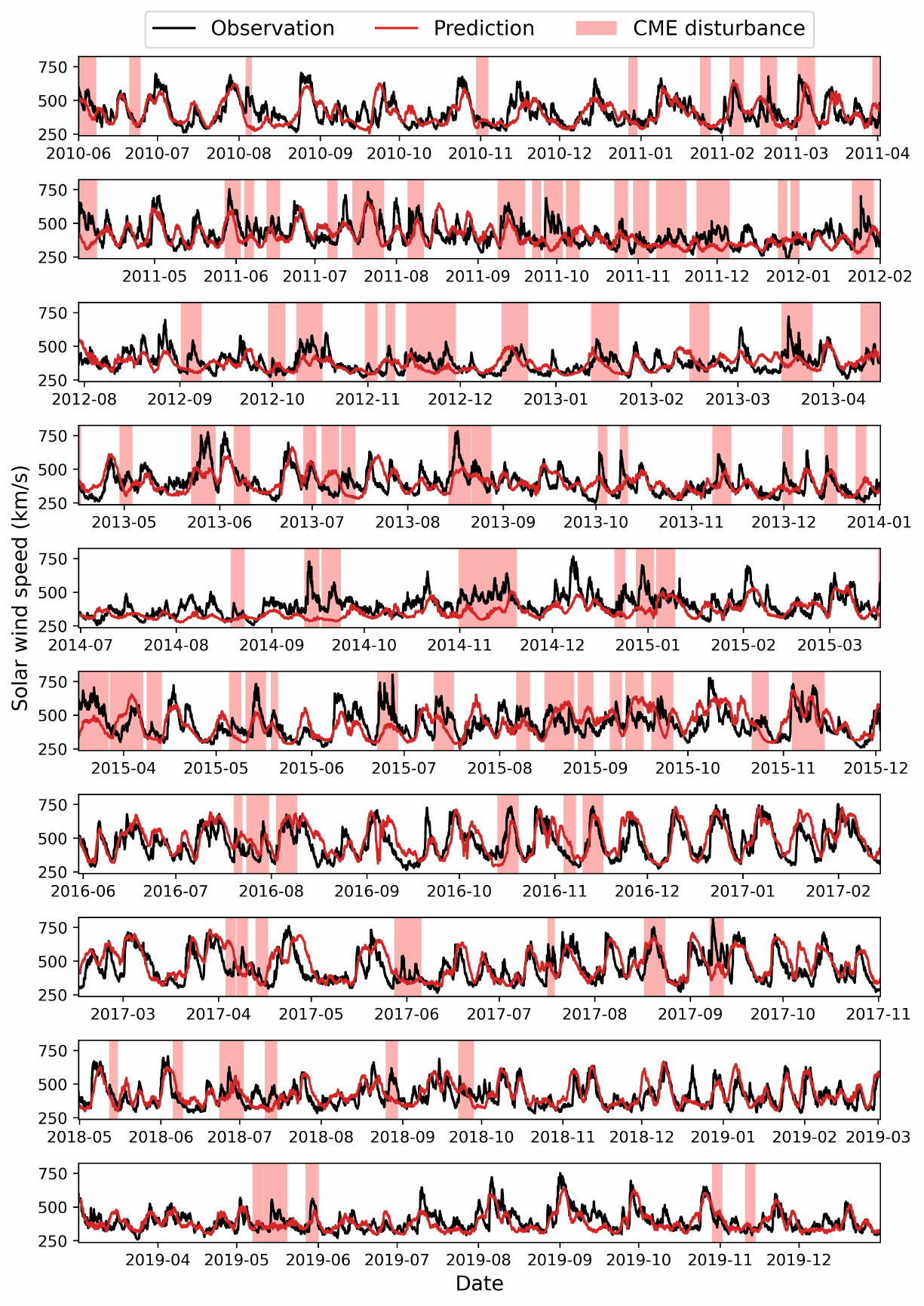}
\caption{Predicted time series of the 4$\times$3 grid model using the distribution transformation on all test data. Excluded CME disturbance intervals are marked red.}
\label{fig:ts}
\end{figure}

\begin{figure}
\centering
\includegraphics[width=\textwidth]{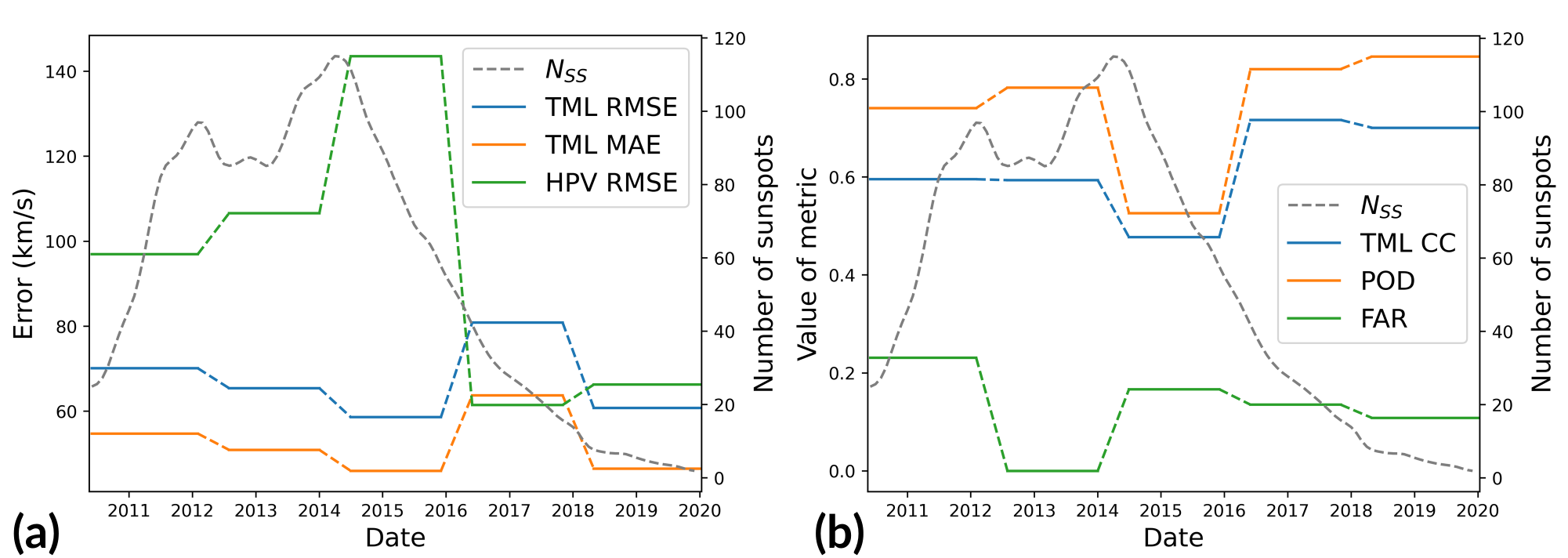}
\caption{The development of different metrics computed for the 4$\times$3 grid model on the CV test sets through the solar cycle. The dashed lines connect the CV test sets. $N_{\rm SS}$ = sunspot number. (a) Error metrics in km/s. (b) Event-based metrics and CC in $[0,1]$. 
TML = Timeline metrics computed for the output of the polynomial model. 
HPV = HSS peak velocity metrics computed after the distribution transformation.}
\label{fig:solar_cycle}
\end{figure}

\begin{figure}
\centering
\includegraphics[width=1.0\textwidth]{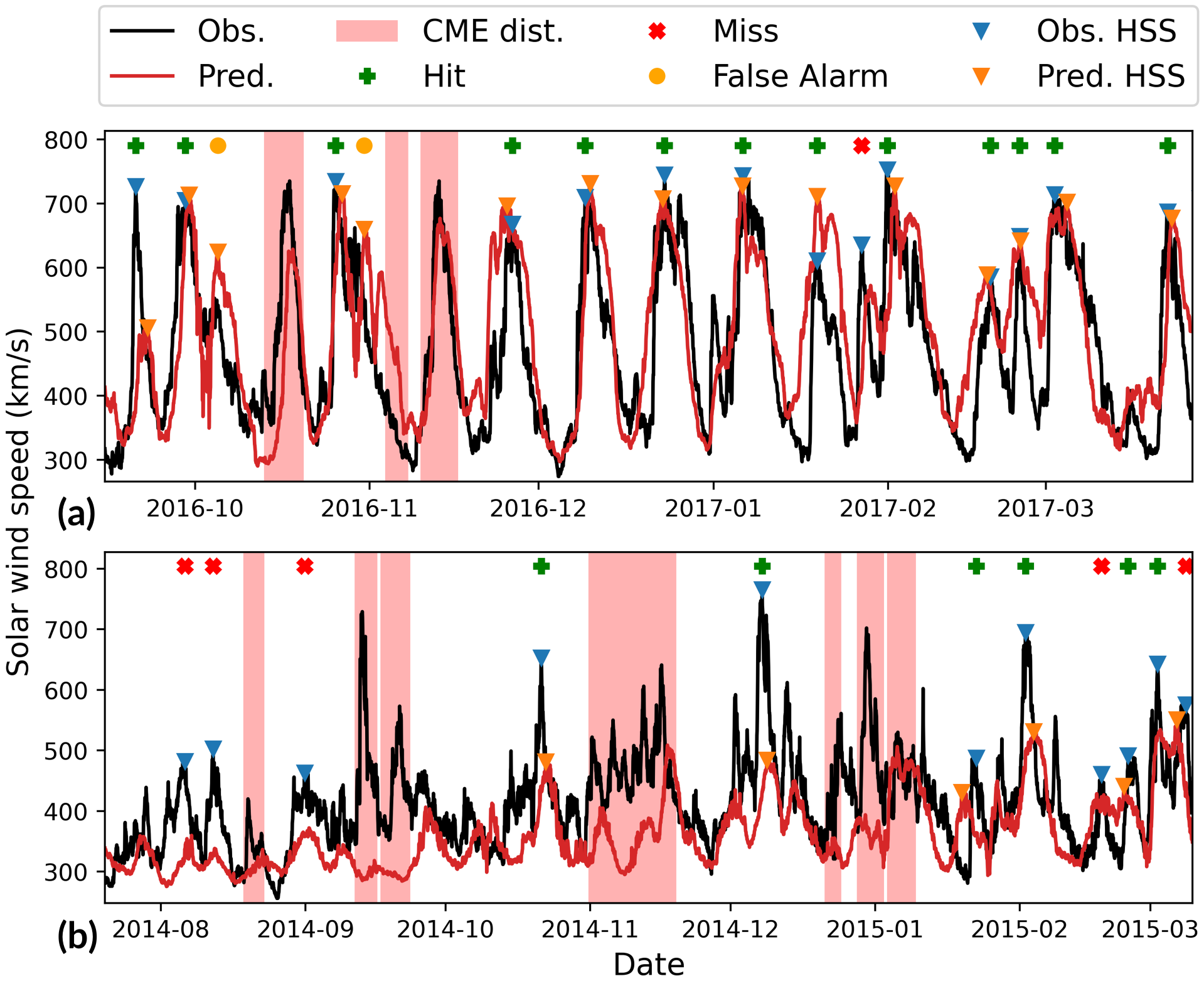}
\caption{Predicted time series of the 4$\times$3 grid model using the distribution transformation. HSS peaks are indicated (triangles) as well as their classification (hit/miss/false alarm). Excluded CME disturbance intervals are marked red. (a) Section from the declining phase of the solar cycle. (b) Section from shortly after the solar maximum.}
\label{fig:zoom}
\end{figure}

Next, we analyze the variability of the model through the solar cycle. 
Figure \ref{fig:ts} shows the time series predicted by all individual CV models on their respective test sets, using the 4$\times$3 grid model. The test data covers almost the entire solar cycle 24. We see that the model generally predicts well the fluctuations of the SWS.
Figure \ref{fig:solar_cycle} shows the corresponding performance metrics. 

In the rising phase of the solar cycle, before 2014, HSS peak velocities are predicted with an intermediate RMSE, relative to the other phases. The POD is high, but the FAR fluctuates between its highest and lowest value. At the end of the rising phase, the FAR even reaches 0. This is because the occurrence rate of fast HSSs decreases with rising solar activity, consequently triggering fewer false alarms. The timeline metrics indicate a high accuracy of the predictions, as the timeline RMSE and MAE are generally low and the timeline CC high.

Around solar maximum, in 2014 and 2015, we observe the worst HSS predictions. The peak velocity RMSE is more than twice as high as during the declining phase, and the POD drops to its minimum. The FAR is slightly higher than in the declining phase. The timeline CC decreases to its minimum, but counterintuitively, the timeline RMSE and MAE achieve their best value. As there are many CME disturbances around solar maximum, which are excluded from the data, the timeline errors are mainly computed for predictions and observations of small fluctuations around the mean, naturally leading to lower errors.

During the declining phase of the solar cycle, from 2016 onward, HSSs are predicted best, as the HSS peak velocity RMSE reaches its minimum and POD its maximum, and the FAR decreases towards the solar minimum.
Also, the timeline CC attains its maximum in the declining phase, indicating that the SWS variations are predicted well. Interestingly, the timeline RMSE and MAE show the opposite behavior, as they increase to their maximum from 2016 to 2017, and only decrease afterward. 
The reason is that many HSSs with large velocity amplitudes, happening very regularly in the declining phase, are not predicted perfectly in time. This temporal error of HSS predictions is caused by our fixed prediction horizon and slightly increases the timeline errors. Thus, in this solar cycle phase, we observe again the trade-off between the timeline performance and the HSS peak performance.

To visualize the solar cycle behavior of our forecast model, we compare two time series segments, one recorded in the declining phase of the solar cycle, where our model performs at its best, and one shortly after the solar maximum, where the performance is at its worst, in Figure \ref{fig:zoom}. In the first time series, showing late 2016 and early 2017, SWS variations and HSSs are very well predicted. The model misses only one HSS that occurs as a SWS increase is predicted too low, and there are only two false alarms.
Even during CME disturbance periods, the SWS predictions often remain good, since many weaker CMEs travel roughly with the ambient solar wind speed.
However, in the second time series, showing late 2014 and early 2015, the speed profiles of HSSs are predicted rather badly, leading to multiple HSS events not being detected and peak values being strongly underestimated. A reason for that is that we train our model on HSSs in other parts of the solar cycle, not reflecting well the properties of HSSs during the solar maximum.

We conclude that our model generally performs well, with its best performance during the declining phase, and also very reliably in the rising phase. During solar maximum, its HSS peak predictions are worse, but the timeline predictions are still comparably accurate.
We find again a trade-off between the timeline and the HSS peak velocity prediction accuracies and observe that a large timeline RMSE does not necessarily mean that the predictions are qualitatively bad.

\section{Comparison to other Models}
\label{sec:comparison}

In this section, we compare our 4$\times$3 grid SWS timeline prediction model to other SWS prediction models. The results are shown in Table \ref{tab:cont_metrics_others}.

\begin{figure}
\centering
\includegraphics[width=1.0\textwidth]{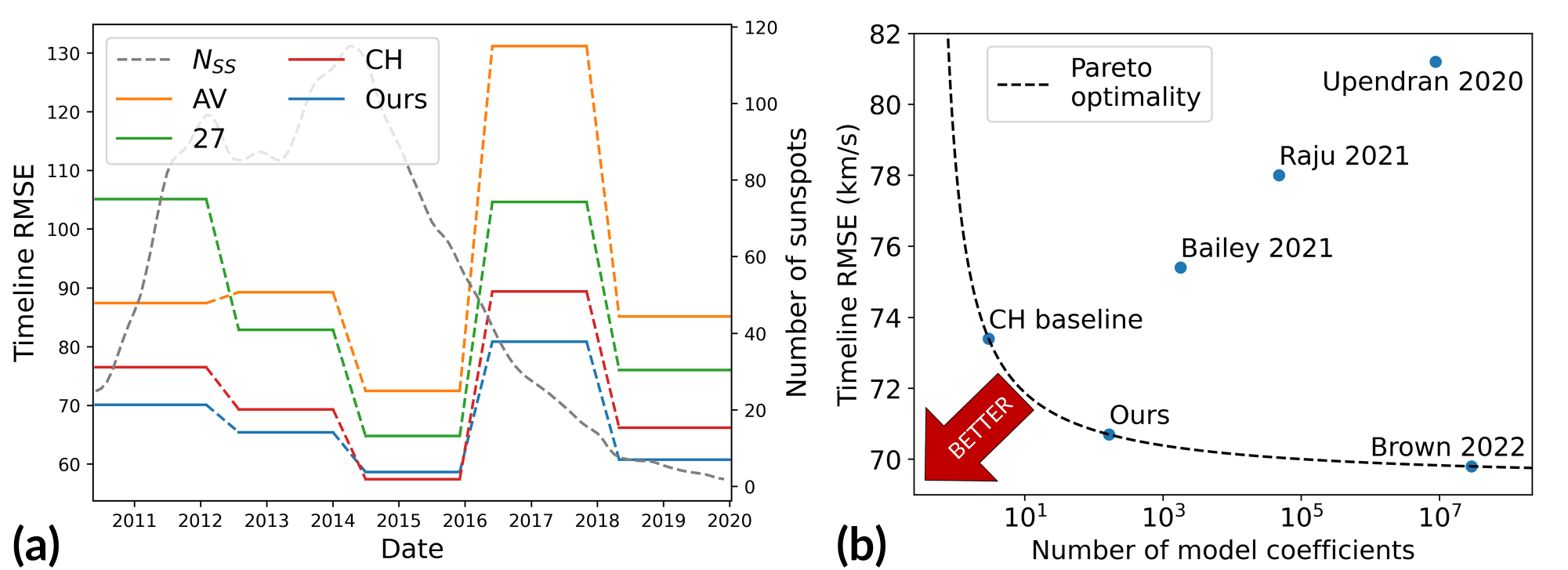}
\caption{Comparison of our model to other models. 
(a) The development of the timeline RMSE computed for the baseline models and the 4$\times$3 grid polynomial model (Ours) on the CV test sets through the solar cycle. The dashed lines connect the CV test sets. $N_{\rm SS}$ = sunspot number. AV = average prediction. 27 = 27-day persistence. CH = coronal hole baseline.
(b) Comparison of the timeline RMSE and the model complexity in terms of the number of learnable model coefficients for different models from the literature and baseline models. The dashed line is the Pareto front (see \ref{app:pareto}).}
\label{fig:comp}
\end{figure}

First, we benchmark against several baseline models. Those are the average prediction model, which predicts for every time point the average SWS of the full training data set, and the 27-day persistence model of \citeA{Owens13}. Further, we design a linear regression model with just two input variables: The SWS measured 27 days ago and the coronal hole area observed four days ago in the two central cells of the 4$\times$3 grid (compare to Figure \ref{fig:preprocessing}), i.e., between 45 degrees northern and southern latitude around the solar equator and between 30 degrees western and eastern longitude around the central meridian. It represents a very simplistic version of our approach, and we call it the coronal hole baseline model.
Figure \ref{fig:comp}a shows the timeline RMSE of our model and the baseline models over the five CV test sets, 
and Table \ref{tab:cont_metrics_others} compares the full set of metrics. Our model clearly outperforms all baseline models.  We perform significantly better than the average prediction model, particularly in the declining phase of the solar cycle. That aligns with the results from Section \ref{sec:solar_cycle}, as it shows that our model is well-adapted to time periods dominated by HSSs. As the solar activity increases, particularly during the solar maximum, the average prediction model becomes relatively better. 
The 27-day persistence model and the coronal hole baseline model follow the same trend as our model, performing better than the average prediction model. The 27-day persistence model generates smaller HSS peak velocity errors because it naturally predicts the same velocity distribution as observed. Therefore, it does not underestimate the peak velocities. The coronal hole baseline model adds a simplistic coronal hole area estimation to the 27-day persistence, which provides a benefit through the entire solar cycle.  Our model resolves the coronal hole area and location by adding a more detailed grid, which provides a further advantage to that. These benefits are largest during the rising and the declining phase of the solar cycle, and least during solar maximum. When solar activity is at its highest, i.e., at solar maximum, the coronal hole baseline model is even competitive with our model.
This shows that other physical factors that are not included in our model may play an increased role in solar maximum.

\begin{table}
\caption{\label{tab:cont_metrics_others}Comparison of other models to our model. The evaluation metrics are computed on the intersection of available dates for each model. We use the predictions of the 4$\times$3 polynomial model for the timeline metrics and apply the distribution transformation for the HSS metrics. RMSE and MAE are given in km/s. Bold numbers indicate the best values.}
\resizebox{\columnwidth}{!}{%
\begin{tabular}{l|lll|lll|llll}
\hline
 & \multicolumn{3}{c|}{Timeline} & \multicolumn{3}{c|}{HSS peak velocities} & \multicolumn{4}{c}{HSS events}\\
\hline
Model  &RMSE   & MAE &   CC   & RMSE  & MAE & CC     & POD           & FAR           & TS    & BS  \\
 \hline
AV        & 97.4 & 76.2 & -0.30 & -     & -     & -    & -    & -    & -    & -    \\
27        & 89.1 & 65.4 & 0.56  & \textbf{78.4}  & \textbf{59.7}  & 0.54 & 0.71 & 0.26 & 0.57 & \textbf{0.95} \\
CH        & 71.2 & 54.8 & 0.67  & 123.4 & 102.8 & 0.52 & 0.69 & \textbf{0.11} & 0.64 & 0.78 \\
Ours      & \textbf{68.1} & \textbf{52.6} & \textbf{0.70}  & 87.5  & 66.8  & \textbf{0.62} & \textbf{0.77} & 0.13 & \textbf{0.69} & 0.88 \\
\hline 
WSA       & 86.5 & 64.1 & 0.38 & 125.5 & 100.7 & 0.10 & 0.57 & 0.17 & 0.51 & 0.68 \\
ESWF      & 91.8 & 69.5 & 0.43 & \textbf{104.6} & \textbf{81.6}  & 0.23 & \textbf{0.68} & 0.35 & 0.50 & \textbf{1.05} \\
Ours      & \textbf{61.8} & \textbf{47.4} & \textbf{0.55} & 121.4 & 101.4 & \textbf{0.34} & \textbf{0.68} & \textbf{0.12} & \textbf{0.63} & 0.77 \\
  \hline
EUHFORIA  & 90.8 & 68.8 & 0.36 & 105.1 & 87.4  & 0.28 & 0.64 & 0.30 & 0.50 & 0.91 \\
Ours      & \textbf{62.9} & \textbf{48.0} & \textbf{0.68} & \textbf{68.6}  & \textbf{53.8}  & \textbf{0.71} & \textbf{0.88} & \textbf{0.12} & \textbf{0.78} & \textbf{1.00} \\
 \hline
GradBoost & 75.6 & 59.1 & 0.65 & 106.6 & 84.1  & \textbf{0.60} & 0.67 & 0.16 & 0.60 & 0.80 \\
Ours      & \textbf{71.2} & \textbf{55.3} & \textbf{0.70} & \textbf{95.4}  & \textbf{73.2}  & 0.59 & \textbf{0.74} & \textbf{0.14} & \textbf{0.66} & \textbf{0.86} \\
 \hline
SwinTrans & \textbf{68.5} & \textbf{52.2} & \textbf{0.71} & 97.3  & 80.9  & \textbf{0.71} & 0.67 & 0.25 & 0.55 & 0.90 \\
Ours      & 69.8 & 54.1 & 0.70 & \textbf{86.6}  & \textbf{68.3}  & 0.63 & \textbf{0.79} & \textbf{0.17} & \textbf{0.68} & \textbf{0.96} \\
\hline
\end{tabular}
}
\end{table}

Next, we compare our model to the predictions of two empirical operational models, the WSA and ESWF model, for the years 2011 to 2014 \cite{Reiss16}. The WSA model derives the coronal magnetic field by magnetic field extrapolation and uses an empirical relation between open magnetic field lines and the SWS \cite{Arge00,Arge03}. The ESWF model uses the empirical relationship between the coronal hole area and the SWS and adapts to the HSS amplitudes of the last three solar rotations \cite{Rotter15}. For all comparisons, we interpolate all predictions to hourly values. Then, we recompute the prediction metrics for the overlapping time interval between the predictions of the comparison model and ours. The comparison to the WSA model shows that we outperform it in all categories, especially for timeline predictions. For HSS peak velocity predictions, the WSA model is just slightly worse, and for HSS events, our model has a significantly higher POD, also resulting in a higher TS. The comparison of our model to the ESWF model is particularly interesting because the latter uses similar inputs as our model, i.e., the coronal hole area in certain solar surface regions. For the accuracy of timeline predictions, our model yields a substantial advantage. For HSS peak velocity predictions, the ESWF model is more accurate.  For HSS events, the ESWF model has the same detection rate as our model but triggers more false alarms. We note that ESWF is the only model in our comparison that does not systematically underestimate the peak velocities. This is because it is not fitted using the standard least squares approach but by adapting the coefficients to the HSS amplitudes. This improves the HSS peak velocity predictions, but also leads to much higher timeline errors, again confirming the trade-off between timeline and peak performance. Our distribution transformation balances this trade-off and keeps the timeline predictions relatively accurate.

Furthermore, we compare our model to the predictions of the physical simulation model EUHFORIA from November 2017 to September 2019 \cite{Hinterreiter19,Samara22}. EUHFORIA models the coronal topology based on photospheric magnetic field observations, derives from that the properties of the young solar wind close to the Sun using a modified empirical WSA model, and propagates this young solar wind to 1 AU using MHD. We observe that the physically simulated EUHFORIA predictions are clearly inferior to our machine learning predictions. We outperform EUHFORIA in all metrics by a large margin. This is likely since it uses static empirical relationships to set the properties of the young solar wind, i.e., it does not adapt well to the current state of the Sun.

Following, we compare our model to two machine learning models, the Swin Transformer of \citeA{Brown22} for 2010 to 2018 and the gradient boosting approach of \citeA{Bailey21} for 2010 to 2017. 
The Swin Transformer is an advanced artificial neural network, taking solar images directly as an input, whereas the gradient boosting algorithm uses magnetic model properties as input.
When compared to the Swin Transformer, our timeline predictions are competitive. On the other hand, for HSS peaks and event detection, we surpass the Swin Transformer in all metrics besides the peak velocity CC. When compared to the gradient boosting model, we find that our model is better by just a small margin, but consistently in all categories besides the HSS peak velocity CC.

Finally, we compare our model to other machine learning models in terms of model complexity, which is measured by the number of model coefficients. For that, we add the models of \citeA{Upendran20} and \citeA{Raju21}, which both use a convolutional neural network based on solar images to predict the SWS. As their predictions are not publicly available, we use the metrics from their publications, which they report with a timeline RMSE of 81.2 km/s and 78.0 km/s, respectively. The model of \citeA{Yang18} is not considered, because they have evaluated their model on the training dataset, i.e., they overpredict their performance. They report an RMSE of 68.0 km/s. Further, we add the coronal hole baseline model for comparison. Figure \ref{fig:comp}b shows the RMSE of the predictions versus the model complexity. 
Our model has the best trade-off between performance and model complexity. With 165 model coefficients based on eight features we achieve an RMSE of 70.7 km/s, outperforming deep neural network approaches with up to 8.8 million coefficients. The only model with a slightly better RMSE is the Swin Transformer of \citeA{Brown22} with 29 million parameters, achieving an RMSE of 69.8 km/s. We conclude that our model achieves a very good trade-off between simplicity and performance. Additionally, we show that accurate SWS predictions are possible with simple models, based on a small number of physical parameters and polynomial functions.

\section{Operational Forecast}
\label{sec:operational}

\begin{figure}
\centering
\includegraphics[width=1.0\textwidth]{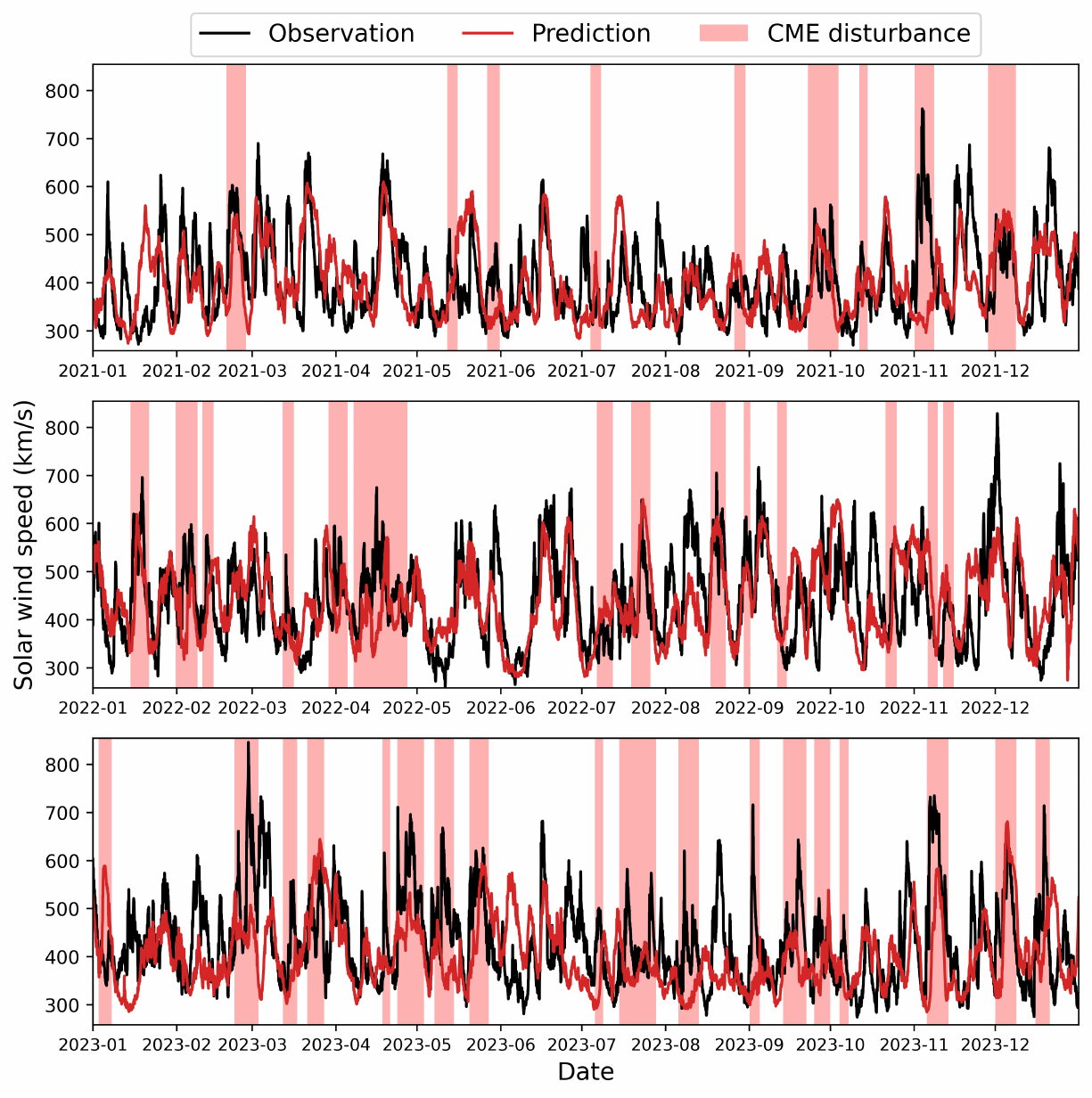}
\caption{Predicted time series of the 4$\times$3 grid model in an operational setting on the current solar cycle 25.}
\label{fig:sc25}
\end{figure}

Finally, we test our approach in an operational setting on data that was excluded from model development and all other experiments in this work. For that, we use the data yet unseen by our model data from 2020 to 2023, comprising the rising phase of solar cycle 25 for the evaluation. These operational predictions are presented in Figure \ref{fig:sc25}.

First, we train an operational 4$\times$3 grid model on the full dataset of solar cycle 24, i.e., on the years 2010 to 2019, and evaluate it on solar cycle 25. We observe that it results in qualitatively inferior HSS peak predictions for solar cycle 25 compared to solar cycle 24. The timeline RMSE is 74.5 km/s for solar cycle 25 as compared to 68.1 km/s for solar cycle 24, and the HSS peak velocity predictions deteriorate to an RMSE of 131.6 km/s for solar cycle 25 as compared to 87.5 km/s from solar cycle 24. Since we observed in Section \ref{sec:solar_cycle} that the model's performance depends on the phase of the solar cycle, we adjust the model training to the correct solar cycle phase. This time, we train the model on the years 2010 to 2015, i.e., the rising phase and solar maximum of solar cycle 24, which is in better accordance with the evaluation dataset that comprises the rising phase and solar maximum of solar cycle 25.  In this model, the timeline RMSE is 76.4 km/s and the HSS peak velocity predictions improve significantly to an RMSE of 110.1 km/s. This shows that the relationship between our physical input parameters and the predicted HSS peak velocities changes with the phase of the solar cycle.

In addition, we believe that solar cycles 24 and 25 also have more fundamental differences. When comparing the linear contribution of coronal hole areas to the prediction of the solar wind speed in solar cycle 25, we find that this linear contribution is significantly lower than in solar cycle 24. To better adapt to solar cycle 25, we use the polynomial model fitted to the rising phase and solar maximum of solar cycle 24 but refit the distribution transformation on the first year of solar cycle 25, i.e., the year 2020. Then, we evaluate our model on the years 2021 to 2023. The results are significantly better than the previous ones due to the adjusted distribution transformation. With this model, we achieve a timeline RMSE of 80.3 km/s, timeline MAE of 60.7 km/s, and a timeline CC of 0.49. These values are slightly worse than the ones of our model in solar cycle 24 but still better than the timeline metrics of other operational forecasts in the rising phase of solar cycle 24 (compare to  Section \ref{sec:comparison}). For the HSS peak velocities, we achieve an RMSE of 92.2 km/s, an MAE of 75.7 km/s, and a CC of 0.49. These numbers are better in the current solar cycle than in the rising phase of the previous one and outperform the other operational models. For HSS events, we obtain a POD of 0.63, FAR of 0.21, TS of 0.54, and BS of 0.8.

We conclude that our model can deliver reliable and robust results in an operational setting on a new solar cycle. Although the timeline performance of our model slightly decreases, HSS predictions remain very accurate, and its performance is superior to that of other operational forecasts in a similar solar cycle phase. In addition, we find that the relationship between the physical input parameters and the HSS peak velocities changes with the phase of the solar cycle and also changes between solar cycles. Therefore, it can be advantageous to recalibrate operational models in regular intervals to adapt to the changing physical conditions.

\section{Discussion}

Our analysis has identified some key insights into the problem of SWS prediction. The primary observation is the deficiency of our model to learn the observed distribution of SWS. This issue is shared by many other machine learning models, e.g., \citeA{Bailey21}, \citeA{Upendran20}, \citeA{Raju21}, \citeA{Brown22}, and can be attributed to the mean squared error (MSE) being employed as loss function to fit the model coefficients. The MSE is very sensitive to outliers. It incentivizes the model to avoid large deviations from the mean to minimize the error during the more frequent quiet times, leading to a bias against predicting high speeds. 
In addition, most HSS predictions are slightly shifted in time. This leads to a large MSE for all data points where predicted and observed geomagnetic storms do not overlap. Lowering the predicted peak decreases the contribution to the MSE during these intervals more than it increases the MSE contribution which happens due to the underestimation of the HSS peak velocity. 
Hence, training a model with the MSE loss function leads to a systematic bias in the predicted distribution and thereby to an underestimation of the peak velocities of HSSs.  
That also explains the trade-off between improving HSS peak velocity accuracy and the timeline RMSE we consistently observe throughout our experiments.
Predicting larger deviations from the mean improves the accuracy for extreme events but is discouraged from the perspective of the timeline RMSE. Notably, this trade-off is also observed for the prediction of other effects, e.g., the Kp index \cite{Shprits19}.

We argue that the focus of SWS prediction models should be on predicting extreme events accurately, as these pose the highest risk for severe space weather conditions, whereas higher errors in quiet times can be tolerated better. Thus, alternatives to the MSE loss functions should be explored. Additionally, we conclude that the RMSE, having the same drawbacks, is not an adequate evaluation measure for space weather predictions, if considered as a standalone metric. It alone should not be used to compare and rank models. We recommend considering metrics that quantify the accuracy of extreme events instead, e.g., the HSS peak velocity RMSE or the POD. 
Our approach shows how that aspect can be emphasized more strongly in a prediction model and how extreme event predictions can be significantly improved. We propose to further explore that direction of research, as there are various advanced methods aiming at approximating the correct distribution while simultaneously minimizing the prediction errors, e.g., distributional regression approaches \cite{Klein23}. 

A further insight is the high variability of the model's performance throughout the solar cycle, caused partly by changing relationships between the input parameters and the SWS.
The variability and cyclic nature of solar activity mean that models require extensive training data to generalize effectively, in the best case spanning several solar cycles. That amount of high-resolution solar images is only slowly becoming available.
Additionally, models should be tested on all phases of the solar cycle, as only then model performance can be robustly evaluated. CV schemes, dividing the data into short segments to train and test on data throughput the complete solar cycle are problematic, because of the necessity to discard long segments of data between training and test sets to account for the 27-day periodic appearance of coronal holes on the solar disk. Otherwise, data leakage caused by long-lasting coronal holes cannot be avoided effectively. That reduces the amount of data that can be used for training even more. On the other hand, if longer segments are chosen for the CV, as we have done, one needs to test on a part of the solar cycle that the model was not trained on. That is another reason for the large variability of model predictions. At the moment, there is no satisfying solution to that problem, but it should be kept in mind when evaluating solar wind models.

Finally, we show that by understanding the physics, it is possible to engineer a small number of physics-based features and to use a simple, not overly expressive prediction model to even surpass the performance of current deep neural network approaches.
Although deep learning models show great potential for prediction tasks, possibly finding new relationships in the data, they need to be fine-tuned and well adapted to their specific task to realize their potential. We exploit the simple relation between the coronal hole area and the SWS to significantly reduce the complexity, increase the interpretability and improve the prediction accuracy of HSS events.

\section{Conclusion}

In this study, we developed a new approach to forecast the SWS originating from coronal holes four days in advance, based on solar images of the current solar rotation and the solar wind speed from the previous solar rotation. We segment coronal holes in solar images and place a grid structure on the segmentation maps to extract the coronal hole area as well as its location on the solar disk. 
We showed that by pairing these features with the SWS of one solar rotation ago, the SWS speed can be predicted well with a simple polynomial regression model. Depending on the grid resolution, prediction models can be constructed that either minimize the prediction error over the entire time series or for the HSS peak velocities. Additionally, we fitted a distribution transformation that scales the SWS predictions to the observed distribution. 
We evaluated and analyzed our approach and found the following main results:
\begin{enumerate}
    \item We achieve an RMSE of 68.1 km/s, a CC of 0.7, and an HSS peak velocity RMSE of 76.8 km/s on an almost 10-year long dataset.
    \item We outperform other models from the literature, including sophisticated deep learning models, MHD models, and empirical SWS forecasts.
    \item We increase the interpretability compared to previous machine learning approaches, reconstructing the SWS variations well using only the coronal hole area, its location, the SWS from the previous solar rotation, and the sunspot number.
     \item Without considering the distribution of SWS, the peak velocities of HSSs are systematically underestimated, because the MSE loss function incentivizes models to avoid predicting large deviations from the mean. 
    This bias can be corrected by a distribution transformation.
    \item The relationship between physical input parameters and the SWS changes over the solar cycle, causing a varying performance of the model, with its best performance in the declining phase of the solar cycle and its worst performance during solar maximum.
    \item We tested our model in an operational setting for the rising phase of solar cycle 25, and achieved an RMSE of 80.3 km/s and an HSS peak velocity of RMSE 92.2 km/s.
\end{enumerate}
For future studies, we recommend
(1) focusing more on the prediction and evaluation of extreme events,
(2) adapting models specifically to the phases of the solar cycle,
(3) modifying model fitting to better approximate the underlying distribution, and
(4) fine-tuning models toward simpler solutions.
Exploring these alternative ways to develop and fit prediction models, possibly avoiding the usage of the MSE, could greatly enhance their value for space weather applications. Additionally, incorporating probabilistic forecasting methods may provide a more nuanced understanding of prediction uncertainties, particularly during storm events.

\newpage

\section*{Data Availability Statement}

The SDO AIA solar image can be downloaded at \url{http://jsoc.stanford.edu/ajax/exportdata.html}. The OMNIWeb solar wind measurements \cite{OMNI} are available at \url{https://omniweb.gsfc.nasa.gov/ow.html} and the OMNI\_M spacecraft positions at \url{https://omniweb.gsfc.nasa.gov/coho/}. Further, the monthly sunspot number from the NOAA SWPC can be accessed at \url{https://www.swpc.noaa.gov/products/solar-cycle-progression}. The Richardson \& Cane ICME list \cite{RCdata} is available at  \url{https://izw1.caltech.edu/ACE/ASC/DATA/level3/icmetable2.htm}. The coronal hole segmentation maps, our machine learning datasets, as well as our compiled HSS and CME lists were published \cite{GFZdata} at \url{https://doi.org/10.5880/GFZ.2.7.2024.001}. The Python code used for the model and the experiments as well as the model predictions are supplied \cite{zenodo} at \url{https://doi.org/10.5281/zenodo.14501728}. The publicly available predictions from \citeA{Brown22} can be accessed at \url{https://github.com/eddbrown/solar-swin-transformer-output-data} and the ones from \citeA{Reiss16} at \url{https://bitbucket.org/reissmar/solar-wind-forecast-verification/src/master/}.


\acknowledgments
We would like to thank Evangelia Samara and Martin A. Reiss for kindly providing us with their prediction data, which made the detailed model comparison possible.
This research is supported by the Helmholtz Imaging Platform, Solar Image-based Modeling (SIM) ZT-I-PF4-016.
Daniel Collin acknowledges the support of the Helmholtz Einstein International Berlin Research School in Data Science (HEIBRiDS). Stefan J. Hofmeister acknowledges support from the NASA Living with a Star Grant 80NSSC20K0183. 


\appendix

\section{Grid Resolution}
\label{app:grid}
In this section, we provide the full overview of the performance of all grid resolutions that we tested. We compare multiple metrics, the number of features selected and used as input for the polynomial model, and give the Lasso penalization hyperparameters of the feature selection and the polynomial model. In Table A1, the models minimizing the timeline RMSE are shown, and in Table A2, the models minimizing the HSS peak velocity RMSE.  

\begin{table}
\centering
\caption{Comparison of a selection of evaluation metrics for models based on different grid resolutions and with hyperparameters optimized to minimize the timeline RMSE. TML = timeline. HPV = HSS peak velocity. n.f. = Number of selected features used as input to the polynomial model. $\gamma_{\rm fs}$ = Lasso penalization hyperparameter of the feature selection model. $\gamma_{\rm pr}$ = Lasso penalization hyperparameter of the polynomial model. Bold numbers indicate the best values.}
\begin{tabular}{llllllllll}
\hline
Grid  & \makecell[cl]{TML\\RMSE}          & \makecell[cl]{TML\\CC}           & \makecell[cl]{HPV\\RMSE}       & POD           & FAR           & TS            & n.f.  & $\gamma_{\rm fs}$ & $\gamma_{\rm pr}$ \\
 \hline
1$\times$1   & 73.6          & 0.63          & 134.6          & 0.60          & 0.09          & 0.56          & \textbf{5} & 3.99$\,\cdot 10^{3}$                        & 1.07$\,\cdot 10^{3}$                        \\
2$\times$1   & 73.6          & 0.63          & 135.4          & 0.60          & 0.09          & 0.56          & \textbf{5} & 2.67$\,\cdot 10^{3}$                        & 1.09$\,\cdot 10^{3}$                        \\
1$\times$3   & 72.1          & 0.65          & 136.6          & 0.65          & \textbf{0.07} & 0.62          & 6          & 3.24$\,\cdot 10^{3}$                        & 1.46$\,\cdot 10^{3}$                        \\
4$\times$3   & 68.1          & 0.70          & 113.3          & \textbf{0.73} & 0.12          & \textbf{0.66} & 8          & 3.46$\,\cdot 10^{3}$                        & 2.94$\,\cdot 10^{5}$                        \\
6$\times$3   & 67.6 & 0.70          & 110.9          & 0.70          & 0.16          & 0.62          & 9          & 2.92$\,\cdot 10^{3}$                        & 2.60$\,\cdot 10^{5}$                        \\
10$\times$3  & 68.2          & \textbf{0.71} & \textbf{110.6} & 0.67          & 0.12          & 0.61          & 19         & 2.34$\,\cdot 10^{3}$                        & 7.69$\,\cdot 10^{5}$                        \\
6$\times$6   & 68.8          & 0.69          & 120.2          & 0.69          & 0.10          & 0.64          & 15         & 2.45$\,\cdot 10^{3}$                        & 3.46$\,\cdot 10^{4}$                        \\
10$\times$6  & 67.7          & 0.70          & 118.9          & 0.66          & 0.08          & 0.63          & 28         & 1.58$\,\cdot 10^{3}$ & 8.35$\,\cdot 10^{4}$ \\
14$\times$6  & \textbf{67.5}          & 0.70          & 116.3          & 0.66          & 0.09          & 0.62          & 38         & 1.93$\,\cdot 10^{3}$                        & 8.34$\,\cdot 10^{4}$                        \\
10$\times$10 & 68.2          & 0.70          & 116.7          & 0.69          & \textbf{0.07} & 0.65          & 42         & 1.48$\,\cdot 10^{3}$ &  6.81$\,\cdot 10^{4}$\\
14$\times$10 & 67.6 & 0.70          & 112.7          & 0.65          & 0.10          & 0.60          & 46         & 1.76$\,\cdot 10^{3}$                        & 4.85$\,\cdot 10^{4}$        \\
\hline
\end{tabular}
\end{table}

\begin{table}
\centering
\caption{Comparison of a selection of evaluation metrics for models based on different grid resolutions and with hyperparameters optimized to minimize the HSS peak velocity RMSE. TML = timeline. HPV = HSS peak velocity. n.f. = Number of selected features used as input to the polynomial model. $\gamma_{\rm fs}$ = Lasso penalization hyperparameter of the feature selection model. $\gamma_{\rm pr}$ = Lasso penalization hyperparameter of the polynomial model. Bold numbers indicate the best values.}
\begin{tabular}{llllllllll}
\hline
Grid  & \makecell[cl]{TML\\RMSE}          & \makecell[cl]{TML\\CC}           & \makecell[cl]{HPV\\RMSE}       & POD           & FAR           & TS            & n.f.  & $\gamma_{\rm fs}$ & $\gamma_{\rm pr}$ \\
 \hline
1$\times$1   & 75.1          & 0.62          & 128.6         & 0.61          & \textbf{0.09} & 0.57          & 8          & 1.31$\,\cdot 10^{3}$                        & 6.04$\,\cdot 10^{4}$                        \\
2$\times$1   & 75.2          & 0.62          & 128.6         & 0.61          & \textbf{0.09} & 0.57          & 9          & 1.55$\,\cdot 10^{3}$                        & 6.01$\,\cdot 10^{4}$                        \\
1$\times$3   & 76.7          & 0.62          & 121.1         & 0.70          & 0.10          & 0.65          & 6          & 4.71$\,\cdot 10^{3}$                        & 1.00$\,\cdot 10^{5}$                        \\
4$\times$3   & 70.5          & 0.67          & 111.3         & 0.71          & 0.13          & 0.65          & \textbf{5} & 9.94$\,\cdot 10^{3}$                        & 1.94$\,\cdot 10^{5}$                        \\
6$\times$3   & 69.2          & \textbf{0.70} & 108.3         & 0.68          & 0.12          & 0.62          & 15         & 1.77$\,\cdot 10^{3}$                        & 4.90$\,\cdot 10^{5}$                        \\
10$\times$3  & \textbf{69.0} & \textbf{0.70} & 104.4         & 0.69          & 0.11          & 0.64          & 21         & 1.71$\,\cdot 10^{3}$                        & 5.28$\,\cdot 10^{5}$                        \\
6$\times$6   & 70.4          & 0.69          & 112.2         & 0.71          & 0.11          & 0.65          & 20         & 1.66$\,\cdot 10^{3}$                        & 1.27$\,\cdot 10^{4}$                        \\
10$\times$6  & 72.8          & 0.67          & 99.5          & 0.70          & 0.13          & 0.64          & 23         & 2.62$\,\cdot 10^{3}$ &  1.76$\,\cdot 10^{5}$ \\
14$\times$6  & 75.3          & 0.62          & 104.6         & 0.75          & 0.15          & 0.66          & 15         & 4.98$\,\cdot 10^{3}$                        & 1.07$\,\cdot 10^{5}$                        \\
10$\times$10 & 72.1          & 0.66          & \textbf{98.9} & \textbf{0.77} & 0.13          & \textbf{0.69} & 18         & 3.98$\,\cdot 10^{3}$ & 1.82$\,\cdot 10^{5}$ \\
14$\times$10 & 70.0          & 0.69          & 99.3          & 0.72          & 0.11          & 0.66          & 42         & 2.05$\,\cdot 10^{3}$                        & 4.72$\,\cdot 10^{5}$     \\
\hline
\end{tabular}
\end{table}

\section{Pareto Optimality}
\label{app:pareto}

In Figures \ref{fig:grid_comp} and \ref{fig:comp}b, where we have given multiple optimization objectives, e.g., the error and the model complexity, a point is called Pareto optimal if one cannot improve one objective without decreasing the others. We can use that concept to generate a Pareto front of all points that are Pareto optimal.
We assume that the error increases rapidly when we reduce the complexity towards zero and that it is possible to decrease the error if we increase the complexity of a model, but with diminishing returns. Thus, we model the Pareto front using the function $f(x)=\frac{a}{x+b}+c$. 
The coefficients $a, b$ and $c$ are obtained by fitting to three points of the Pareto optimal vertices of the convex hull of all points, chosen such that $f(x) \leq p_i$ for all points $p_i$.

\newpage

\bibliography{literature}

\end{document}